\pdfoutput=1

\documentclass[prx,aps,12pt,superscriptaddress]{revtex4-2}
\usepackage{amsmath,amssymb}
\usepackage{bm}
\usepackage{comment}
\usepackage{graphicx}
\usepackage[dvipsnames]{xcolor}
\usepackage{mathrsfs}
\usepackage{amsmath}
\usepackage{amsfonts}
\usepackage{dsfont} 

\newcommand\x{\mathbf{x}}

\newcommand\q{\mathbf{q}}

\newcommand\<\langle
\renewcommand\>\rangle

\renewcommand\d{\partial}

\newcommand{\be}{\begin{equation}}
\newcommand{\ee}{\end{equation}}
\newcommand{\bea}{\begin{eqnarray}}
\newcommand{\eea}{\end{eqnarray}}
\newcommand{\ie}{\begin{equation}\begin{aligned}}
\newcommand{\fe}{\end{aligned}\end{equation}}
\renewcommand\d{\partial}

\usepackage{xcolor}
\usepackage{amsmath}
\DeclareMathOperator{\Tr}{Tr}

\newcommand\+\dagger

\usepackage{comment}
\usepackage[colorlinks=true,citecolor=blue,urlcolor=blue,linkcolor=black,pdfstartview=FitH,bookmarksopen]{hyperref}
\renewcommand\d\partial
\newcommand\vareps{\varepsilon}
\usepackage{booktabs,tabularx}
\usepackage{footnote}

\begin{document}

\title{Chiral Graviton Theory of Fractional Quantum Hall States} 

\author{Yi-Hsien Du}
\email[Electronic address:$~~$]{yhdu@mit.edu}
\affiliation{Department of Physics, Massachusetts Institute of Technology, Massachusetts 02139, USA}
\affiliation{Kadanoff Center for Theoretical Physics, University of Chicago, Chicago, Illinois 60637, USA}

\begin{abstract}
Recent polarized Raman scattering experiments indicate that fractional quantum Hall systems host a chiral spin-2 neutral collective mode, the long-wavelength limit of the magnetoroton, which behaves as a condensed-matter graviton. We present a nonlinear, gauge-invariant effective theory by gauging area-preserving diffeomorphisms (APDs) with a unimodular spatial metric as the gauge field. A Stueckelberg construction introduces an APD-invariant local potential that aligns the dynamical metric with a reference geometry, opening a tunable gap while preserving gauge redundancy. Together with a geometric Maxwell kinetic sector and the Wen–Zee and gravitational Chern–Simons terms, the theory yields a gapped chiral spin-2 excitation consistent with universal long-wavelength constraints. The tunable gap emerges naturally from symmetry and provides a route to an isotropic-nematic quantum critical point where the spin-2 mode softens. We further establish a linear dictionary to quadrupolar deformations in composite Fermi liquid bosonization, and outline applications to fractional Chern insulators as well as higher-dimensional generalizations. Finally, the approach can be extended to non-Abelian fractional quantum Hall states, capturing both spin-2 and spin-$3/2$ neutral modes.
\end{abstract}

\maketitle

\tableofcontents

\section{Introduction}

Recently, it has been proposed that fractional quantum Hall (FQH) systems host a chiral spin-2 neutral collective mode~\cite{haldane2009hallviscosityintrinsicmetric,PhysRevLett.107.116801,Maciejko:2013dia,Golkar:2013gqa,PhysRevX.7.041032,Nguyen:2017qck,PhysRevLett.123.146801}, which is the magnetoroton~\cite{PhysRevB.33.2481} with zero wave number. The spin and chirality of this mode can be probed with circularly polarized Raman scattering~\cite{Golkar:2013gqa,PhysRevLett.123.146801,PhysRevResearch.3.023040}.  In particular, the intensity of the magnetoroton line in Raman scattering experiments does not diminish at zero momentum, and lattice-induced breaking of continuous rotational symmetry down to $C_4$ by is numerically small and does not prevent the determination of the sign ($2$ or $-2$) of the spin~\cite{PhysRevResearch.3.023040}.  Recent polarized Raman scattering experiments have directly confirmed both the spin-2 nature and the chirality of the zero-momentum magnetoroton at filling fractions $\nu=1/3$, $2/5$, $3/5$ and $2/3$~\cite{Liang:2024dbb}.

The spin-2 mode has been argued to be a condensed-matter analog of the graviton: it describes fluctuations of a tensor order parameter.  Haldane~\cite{PhysRevLett.107.116801} suggested that this order parameter matrix is in fact a spatial metric, a fundamental degree of freedom of the FQH liquid.  This perspective is made concrete in bimetric theory~\cite{PhysRevX.7.041032}, where an unimodular metric captures the long-wavelength magnetoroton. The leading time-derivative term is the mixed Chern–Simons term between the electromagnetic $U(1)$ gauge field and the spin connection built from the metric, supplemented by a phenomenological mass for the spin-2 mode. The resulting ``graviton'' is a gapped chiral spin-2 excitation, similar to what has been observed in experiments.  

Beyond conventional Landau-level experimental platforms in semiconductor heterostructures, FQH physics also appears in two-dimensional van der Waals moir\'e materials~\cite{Cai_2023,Park_2023,Zeng_2023,Lu_2024} and in twisted bilayer graphene~\cite{Xie_2021} without an external magnetic field -- a phenomenon called the fractional quantum anomalous Hall (FQAH) effect. In particular, twisted transition metal dichalcogenides (tTMDs) realize flat, topologically nontrivial Chern bands that closely resemble Landau levels. 

The FQAH states emerging at fractional filling of these bands, often called moir\'e fractional Chern insulators (FCIs), are believed to share the same universality class of Landau-level FQH states. Moreover, field-induced FQH states have been observed in related systems where strong periodic superlattice potentials are present~\cite{Spanton_2018,aronson2024displacementfieldcontrolledfractionalchern}. Importantly, optical experiments provide powerful probes of the neutral collective modes in moir\'e FCIs hosted by tTMDs, accessible either directly through terahertz (THz) spectroscopy or indirectly via inelastic Raman scattering~\cite{paul2025shininglightcollectivemodes,kousa2025theorymagnetorotonbandsmoire}; for example, in twisted MoTe$_2$ a chiral graviton is present but not the lowest-energy neutral excitations.

In this work, we systematically develop a nonlinear effective theory for a massive chiral graviton in (2+1) and (3+1) dimensions, emphasizing the structure and realization of the gauge symmetry. We formulate this nonlinear chiral gravity theory on a nonrelativistic spacetime with an absolute time, requiring invariance of the action with respect to spatial, but in general time-dependent, volume-preserving diffeomorphisms (VPDs), treated as a local gauge redundancy. In (2+1) dimensions the gauge fields are a unimodular spatial matrix $g_{ij}$ and a scalar potential $A_0$, reminiscent of unimodular gravity in relativistic contexts~\cite{Einstein:1919gv,Anderson:1971pn}.

A Stueckelberg construction furnishes a local, APD-invariant potential that aligns the dynamical metric with a reference geometry and opens a tunable gap while preserving the gauge redundancy. In unitary gauge the would-be Nambu–Goldstone mode is eaten and the spin-2 excitation becomes massive; the APD symmetry is nonlinearly realized rather than broken. By contrast, the isotropic–nematic transition corresponds to spontaneous breaking of global spatial rotations. Augmenting the Stueckelberg sector with a parity-even geometric Maxwell kinetic term and with Wen–Zee and gravitational Chern–Simons terms yields a chiral, gapped spin-2 mode consistent with universal long-wavelength constraints. At quadratic order the symmetry-based mass reduces to the familiar bimetric form, providing a controlled knob that tunes the system from the isotropic phase to a nematic quantum critical point by closing the spin-2 gap. We also establish a linear dictionary to quadrupolar deformations in composite Fermi liquid bosonization near half filling, thereby connecting the graviton description to Fermi surface dynamics. Finally, we outline extensions to FCIs, where band geometry (Berry curvature and quantum metric) naturally enter, and to higher-dimensional generalizations based on gauged VPDs.

A further motivation is to address non-Abelian FQH phases, notably at $\nu=5/2$~\cite{Nayak_2008}, which host non-Abelian anyon excitations that offer a potential platform for fault-tolerant topological quantum computation~\cite{NAYAK1996529,Nayak_2008,KITAEV20032}. In the Pfaffian state (and its particle–hole conjugate, the anti-Pfaffian state), a neutral spin-$3/2$ fermionic mode accompanies the spin-2 magnetoroton, suggesting a unified supergravity description of both modes. Prior work has explored related ideas for the Haldane-Rezayi state and special cases~\cite{PhysRevLett.125.077601,PhysRevB.107.125119}; here we frame a path toward gapped non-Abelian FQH phases such as Read–Rezayi states~\cite{PhysRevB.59.8084,PhysRevB.60.8827,PhysRevB.81.045323,PhysRevB.81.155302}.

\section{Chiral Graviton Theory in (2+1) dimensions}

\subsection{Gauge symmetry}\label{sec:gauge-symmetry}

As the gauge symmetry, we consider the group of area-preserving diffeomorphisms (APDs). Infinitesimally, these transformations act as $x^i\to x^i+\xi^i$, where the vector field $\xi^i$ is divergence-free, $\d_i\xi^i=0$.  
On a simply connected two-spatial-dimensional manifold, this constraint can be solved in terms of a single scalar gauge parameter $\lambda(x)$:
\begin{equation}
\xi^i(x)=\ell^2\varepsilon^{il}\d_l \lambda(x)~.
\end{equation}
Here $\ell$ is a problem-dependent length scale (the magnetic length in the quantum Hall setting).

We adopt the ``passive'' viewpoint: an infinitesimal APD acts on any function $f(x)$ transforms as $f(x)\to f(x)+ \delta_\lambda f(x)$, where
\begin{equation}
  \delta_\lambda f(x) = - \xi^k \d_k f = -\ell^2\varepsilon^{kl}\d_k f\d_l \lambda \equiv \{ \lambda,\, f\}~,
\end{equation}
i.e., via the classical Poisson bracket $\{f,\, g\} =\ell^2\varepsilon^{kl}\d_k f \d_l g$. Equivalently, we can think of $\delta_\lambda$ as a differential operator: 
\begin{equation}
    \delta_\lambda= - \mathcal{L}_\xi = \ell^2\vareps^{ij}\d_i\lambda\d_j~,
\end{equation}
where $\mathcal L_\xi$ is the Lie derivative along the vector field $\xi^i$ and its negative furnishes the APD generator. These infinitesimal transformations close under commutation and realize an APD (or $w_\infty$) algebra,
\begin{equation}
[\delta_{\lambda_1}, \, \delta_{\lambda_2}] = \delta_{\{\lambda_1,\ \lambda_2\}}~, \qquad
  \{\lambda_1,\, \lambda_2\} = \ell^2\varepsilon^{kl}\d_k\lambda_1\d_l\lambda_2 ~.
\end{equation}

Finite APDs are obtained by exponentiating infinitesimal APDs, i.e.,
\begin{equation}
U(\lambda) = \exp(\ell^2\varepsilon^{kl}\d_k \lambda \d_l)~.
\end{equation}
Clearly, these finite APDs form an infinite-dimensional Lie group, where the multiplication rule follows from the Baker-Campbell-Hausdorff formula:
\begin{equation}
 U(\lambda_1) U(\lambda_2) = U\! \left(\lambda_1 + \lambda_2 + \frac12 \{\lambda_1,\, \lambda_2\} + \cdots \right)~.
\end{equation}

\subsection{Gauge field}\label{sec:gauge-field}

\begin{figure}[t]
  \centering
  \includegraphics[width=0.45\linewidth]{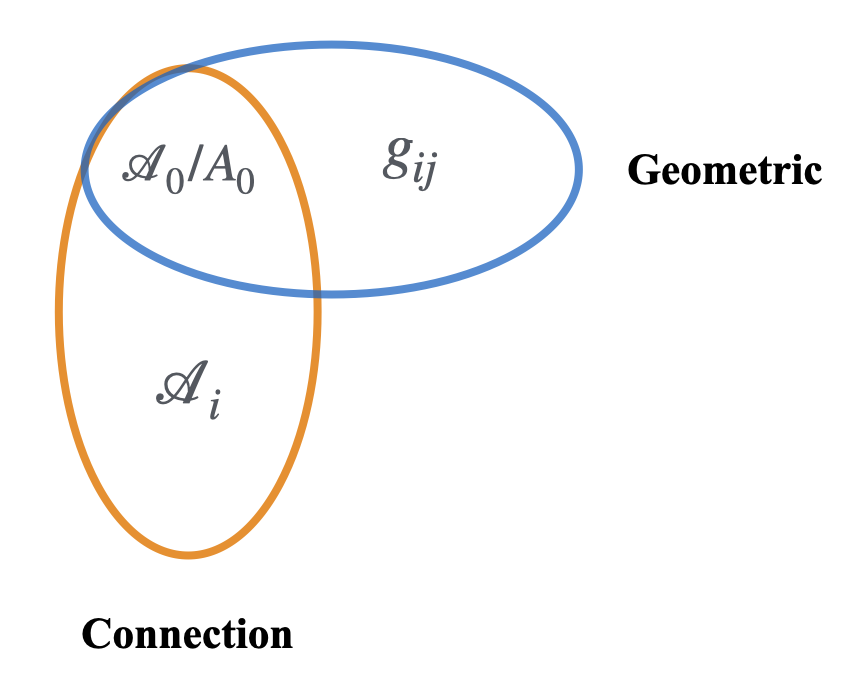}
  \caption{Connection vs.\ geometric realizations of APD.
We employ two equivalent realizations of the APD gauge redundancy.
(i) \emph{Connection realization:} an APD gauge connection
$\mathcal A_\mu=(\mathcal A_0,\mathcal A_i)$.
(ii) \emph{Geometric realization:} a pair $(A_0,g_{ij})$ consisting of an APD temporal potential $A_0$ and a unimodular spatial metric $g_{ij}$.
We identify the temporal components, $A_0\equiv\mathcal A_0$.
A local field redefinition (given later) trades $\mathcal A_i$ for $g_{ij}$, so the two descriptions are equivalent and should not be treated as independent. To avoid confusion, we reserve $\tilde A_\mu$ for the external
  electromagnetic $U(1)$ probe; $A_0$ above denotes the APD temporal potential.}
\label{fig:apd-gauge}
\end{figure}

There are at least two different realizations with APD being the gauge symmetry (Fig.~\ref{fig:apd-gauge}).  

\paragraph*{Connection realization.} The most straightforward way is to introduce a gauge connection $\mathcal A_\mu$ transforming under APD as~\cite{Du_2024}
\begin{equation}\label{apd-gauge}
  \delta_\lambda \mathcal A_\mu = \d_\mu \lambda +\{\lambda ,\, \mathcal A_\mu\}~.
\end{equation}
The gauge theory can be constructed in close analogy with non-Abelian gauge theories; in particular, the field strength is defined by
\begin{equation}
   \mathcal F_{\mu\nu} = \d_\mu \mathcal A_\nu - \d_\nu \mathcal A_\mu - \{\mathcal A_\mu,\, \mathcal A_\nu\}~,
\end{equation}
which transforms homogeneously, $\delta_\lambda \mathcal F_{\mu\nu} = \{\lambda,\, \mathcal F_{\mu\nu} \}$. A Maxwell action
\begin{equation}
   S =  \int_\x \left(\frac{\epsilon_0}2 \mathcal F_{0i}^2 - \frac{1}{2\mu_0} \mathcal F_{12}^2 \right)~,
\end{equation}
is automatically APD invariant, since the Lagrangian density is an APD scalar and $\int_\x\, \{A,\, B\}=0$ for fields vanishing at infinity. Chern–Simons terms may also be constructed~\footnote{In contrast to compact non-Abelian gauge theories, the APD connection lives in a non-compact, infinite-dimensional gauge redundancy. Consequently, ``Chern–Simons'' we mean functionals built with the Poisson bracket: $S_{\rm PCS}
=\frac{k}{4\pi}\!\int_\x\;\varepsilon^{\mu\nu\rho}
\left(
\mathcal A_\mu\,\partial_\nu \mathcal A_\rho
+\frac{1}{3}\,\mathcal A_\mu\{\mathcal A_\nu,\mathcal A_\rho\}\right)$. Because the APD gauge algebra is non-compact and infinite-dimensional, these terms are best viewed as geometric Chern-Simons terms reduce to Wen–Zee and gravitational Chern-Simons after rewriting in metric variables. Because APD is non-compact, these terms are not level-quantized and, by themselves, do not imply genus-dependent ground-state degeneracy; the latter resides in the compact charge sector, e.g., a $K$-matrix Chern-Simons theory.}. A curved-space formulation appears in App.~\ref{sec:apd-curved}.

\paragraph*{Geometric realization.} An alternative, geometric realization of the APD gauge symmetry introduces, the time component $A_0$ and the spatial metric $g_{ij}$ as the gauge fields~\cite{Du_2022}. Their infinitesimal transformations are given as follows (for later convenience, we record several algebraically equivalent expressions of each quantity):
\begin{align}\label{APD-transf}
  \delta_\lambda A_0 & = \dot\lambda - \mathcal L_\xi A_0 = \dot\lambda - \xi^k\d_k A_0 \equiv \dot\lambda + \{\lambda, \, A_0\}~,\\
  \delta_\lambda g_{ij} & = -\mathcal L_\xi g_{ij} = -\xi^k\d_k g_{ij} - g_{kj}\d_i\xi^k - g_{ik}\d_j\xi^k\nonumber\\ 
  & = -\ell^2 \varepsilon^{kl}(\d_k g_{ij}\d_l \lambda + g_{kj}\d_i\d_l\lambda + g_{ik} \d_j\d_l\lambda)~.
\end{align}

Introducing the raising and lowering spatial indices with $g_{ij}$ and its inverse $g^{ij}$, e.g.,
\begin{equation}
   \xi_i = g_{ij} \xi^j~,
\end{equation}
and spatial covariant derivative $\nabla_i$ with Christoffel symbol $\Gamma^k_{ij}$, e.g.,
\begin{equation}
   \nabla_i \xi_j = \d_i \xi_j - \Gamma^k_{ij}\xi_k ~,\qquad
   \Gamma^k_{ij} = \frac12 g^{kl}(\d_i g_{lj}+\d_j g_{li}- \d_l g_{ij})~,
\end{equation}
one may write
\begin{equation}
 \delta_\lambda g_{ij} = - \nabla_i\xi_j - \nabla_j \xi_i~.
\end{equation}
At the linearized level~\cite{gu2006latticebosonicmodelquantum,xu2006novelalgebraicbosonliquid}, expand around flat space as $g_{ij}=\delta_{ij}+h_{ij}+\mathcal{O}(h^2)$. Then an infinitesimal APD acts as
\begin{align}
  \delta A_0 &= \dot \lambda~, \\
  \delta h_{ij} &= -\ell^2 (\vareps^{il}\d_j\d_l\lambda + \vareps^{jl}\d_i\d_l\lambda)~.
\end{align}

\paragraph*{Equivalence and choice of gauge fields.}
We use the connection realization $\mathcal A_\mu$ and the geometric realization $(A_0,g_{ij})$ as two coordinate choices of the same APD gauge redundancy. A local field redefinition (given later) trades $\mathcal A_i$ for the unimodular metric $g_{ij}$, with the temporal components identified, $A_0\equiv\mathcal A_0$, so no additional degrees of freedom are introduced and the two descriptions are not to be used as independent fields. The geometric variables make the physical spin-2 content and coupling to curvature explicit, while the connection variables are often algebraically convenient. 

\paragraph*{Global magnetic translation and rotation.} The trivial background $A_0=0$, $g_{ij}=\delta_{ij}$ is invariant under a finite-dimensional global subgroup of APDs generated by quadratic polynomials
\begin{equation}
  \lambda(\x)=\lambda_0 + a_i x^i + \frac{b}{2} x^2~,
\end{equation}
where $\lambda_0$, $a_i$ and $b$ are constant parameters. The associated APD vector field 
\begin{equation}
  \xi^i(\lambda)=\ell^2\varepsilon^{ij}\partial_j\lambda
  = \ell^2\varepsilon^{ij}a_j+ b \, \ell^2 \varepsilon^{ij}x_j~,
\end{equation}
generates uniform magnetic translations and rotations, respectively. Their Poisson brackets reproduce the expected projective structure,
\begin{equation}
  \{a \cdot x,\, a' \cdot x\}=\ell^2\varepsilon^{ij}a_i a'_j~,\qquad
  \Big\{\frac{x^2}{2},\,a \cdot x\Big\}=\ell^2\varepsilon^{ij}x_i a_j~,
\end{equation}
mirroring the Heisenberg algebra of magnetic translations on the lowest Landau level (LLL).

\paragraph*{``Electric'' and ``magnetic'' field strengths.}We can define the ``drift velocity'' as
\begin{equation}
v^i\equiv \ell^2 \varepsilon^{ij}\partial_j A_0~,\qquad v_i\equiv g_{ij}v^j~,
\end{equation}
which transform under APDs as
\begin{equation}
\begin{split}
\delta_\lambda v^i &= \dot\xi^i -\mathcal L_\xi v^i =\dot\xi^i+\{\lambda,\, v^i\}+v^j\d_j \xi^i~,\\
\delta_\lambda v_i &= g_{ik} \dot \xi^k -\mathcal L_\xi v_i= g_{ik} \dot \xi^k +  \{\lambda,\, v_i\}-v_j\d_i \xi^j~.
\end{split}  
\end{equation}
Under an uniform Galilean boost $\xi^i = c^i t$, the drift field shifts by a constant $\delta_\lambda v^i = c^i$, as appropriate for a velocity. One can ``covariantize'' time derivatives: $\nabla_t = \d_t + \mathcal L_v$. In particular,
\begin{equation}
  \nabla_t g_{ij} = \dot g_{ij} + \nabla_i v_j + \nabla_j v_i ~.
\end{equation}
This motivates defining APD ``electric'' and ``magnetic'' field strengths,
\begin{equation}\label{maxwell}
\ell^2 \left(\vareps_{ik}\mathcal{E}_{jk} + \vareps_{jk}\mathcal{E}_{ik} \right) = \nabla_t g_{ij}~, \qquad \ell^2 \mathcal{B} = \frac{1}{2} R~,
\end{equation}
where $R$ is the spatial Ricci scalar. By construction, $\mathcal{E}_{ij}$ and $\mathcal{B}$ transform covariantly under APDs~\cite{PhysRevB.110.035164,Nguyen_2024}.

\subsection{The effective action}

In incompressible fractional quantum Hall states, such as the Laughlin, Moore-Read, Read-Rezayi, and Jain states, the entire neutral magnetoroton branch is gapped; in particular, the long-wavelength chiral spin-2 ``graviton'' has a finite gap~\cite{PhysRevLett.50.1395,MOORE1991362,PhysRevB.59.8084,PhysRevLett.63.199}. In contrast, at filling fraction $\nu=1/2$ the systems forms a compressible composite Fermi liquid: with unscreened Coulomb interactions the long-wavelength magnetoroton is absent and the spin-2 sector is gapless and Landau-damped, i.e., a ``massless graviton'' without a sharp pole~\cite{PhysRevLett.128.246402}; for short-range interactions the non–Fermi-liquid behavior is even stronger~\cite{Nayak_1994,NAYAK1994534}. These observations motivate a nonlinear effective theory for a gapped spin-2 mode in incompressible phases. 

Our starting point is a geometric Maxwell–Chern–Simons structure. The Maxwell term supplies the parity-even kinetic energy for the propagating mode, while the Wen–Zee and gravitational Chern–Simons terms encode the parity-odd Berry structure; together they produce a gapped chiral spin-2 excitation. In our non-compact APD formulation these geometric Chern-Simons terms do not by themselves constitute a standalone topological field theory and therefore do not exhibit genus-dependent ground-state degeneracy~\cite{PhysRevB.40.7387,PhysRevB.41.9377}; the latter belongs to the compact charge sector (e.g., a $K$-matrix $U(1)^r$ Chern–Simons theory for an internal gauge field $a_\mu$), which we keep implicit and which couples to the neutral spin-2 sector through background geometry.

Collecting the terms in~\eqref{maxwell}, a geometric Maxwell formulation in (2+1) dimensions is
\begin{equation}
  \mathcal{L}_\text{Maxwell} = \frac{c_1}4 (\nabla_{\!t} g_{ij})^2 -c_2 R^2~,
\end{equation}
with constants $c_1$ and $c_2$. By contrast, a linear Einstein–Hilbert term integrates to the topological Euler characteristic, $\chi=\tfrac{1}{4\pi}\int_\x R$, in two spatial dimensions, it does not contribute to the bulk dynamics and can be neglected from the action.

\paragraph*{Power counting and renormalizability.}A natural question arises: is this nonrelativistic unimodular gravity theory renormalizable? Although a full loop analysis can be carried out, engineering power counting already suffices. In the ultraviolet (UV), the spin-2 dispersion $\omega\sim q^2$ fixes the dynamical critical exponent to $z=2$ and scaling dimensions
\begin{equation}
[x] = -1~, \qquad [t] = -2~, \qquad [g] = 0~,
\end{equation}
so all couplings in $\mathcal L$ are nonnegative-dimension, indicating that the theory is renormalizable through power counting~\cite{FRADKIN1982469,Fradkin1984,PhysRevD.16.953,Deser_1990}. This mirrors Horava-Lifshitz–type analyses~\cite{Ho_ava_2009}, where in $(d+1)$ dimensions the Lifshitz (anisotropic) scaling assignment $[x]=-1,\ [t]=-z$ renders scalar fields dimensionless at $z=d$. In the infrared (IR), if the dispersion crosses over to $\omega\sim q^4$ one has
\begin{equation}
  [x]=-1~,\qquad [t]=-4~,\qquad [g]=1~,
\end{equation}
so the metric acquires positive mass dimension.

\subsection{Chern-Simons terms}

An important class of topological contributions to the effective field theory is the Chern-Simons terms. In $(2+1)$ dimensions, the FQH effect is fruitfully described by effective actions that capture the bulk and boundary dynamics of fermionic liquids. On manifolds with boundary the bulk Chern–Simons action is not gauge invariant, but gauge invariance of the full theory is restored by anomaly inflow from chiral edge modes. The latter are described by a chiral boson with a Floreanini–Jackiw action~\cite{PhysRevB.41.12838,1992IJMPB...6.1711W,Floreanini:1987as}. More generally, in higher dimensions, specifically in $4k+2$ spacetime dimensions, chiral (self-dual) fields generate gravitational anomalies~\cite{PhysRevLett.63.728}.

To construct the geometric Chern-Simons term, we introduce the spin connection
\begin{equation}
\omega_\mu = \frac12 \vareps^{ab} e^{a\nu}\nabla_\mu e^b_\nu ~,
\end{equation}
where $e^a_\nu$ represents the vielbein, with $a = 1,2$, and relates to the metric via $g^{\mu\nu} = e^{a\mu} e^{a\nu}$. The APD transfromation on the vielbein follows from that on the metric~\eqref{APD-transf} is
\begin{equation}
    \delta_\lambda e^a_\nu= -\mathcal L_\xi e^a_\nu = -\xi^i \d_i e^a_\nu - \d_\nu \xi^i e^a_i ~.
\end{equation} Under local $O(2)$ rotations of the vielbeins:
\begin{equation}
e^a(x) \to e^a(x) + \alpha(x) \vareps^{ab} e^b(x)~,
\end{equation}
the spin connection transforms as an Abelian gauge field 
\begin{equation}
    \omega_\mu \to \omega_\mu +\d_\mu \alpha~.
\end{equation}
Using $e^{aj}\equiv e^a_i g^{ij}$, the components can be written explicitly as
\begin{equation}
\begin{split}
\omega_0 &=\frac12 (\vareps^{ab} e^{aj}\d_0 e^b_j
  +\vareps^{ij} \d_i v_j)~,\\
\omega_i &=\frac12 (\vareps^{ab} e^{aj}\d_i e^b_j-\vareps^{jk}\d_jg_{ik}) ~.
\end{split}
\end{equation}
The additional $1/2 \, \vareps^{ij} \d_i v_j$ term in $\omega_0$ ensures that $\omega_\mu$ transforms as a 1-form under spatial diffeomorphism. In two spatial dimensions, the Ricci scalar in terms of the spin connection is defined by
\begin{equation}
R= 2 \, \vareps^{ij} \d_i \omega_j~.
\end{equation}

\paragraph{Wen-Zee term.}A key Chern-Simons term contribution to the effective action of the fractional quantum Hall effect is the Wen-Zee term~\cite{PhysRevLett.69.953,Du_2022},
\begin{equation}
\mathcal{L}_{\text{WZ}} = \frac\kappa{4\pi}\vareps^{\mu\nu\lambda}\omega_\mu \d_\nu A_\lambda = \frac\kappa{4\pi}\left(\frac12 A_0 R + \frac{\omega_0}{\ell^2} \right)~,
\end{equation}
where we have set a static, uniform condition with $\dot A_i = 0$ and $\varepsilon^{ij} \partial_i A_j = \ell^{-2}$. The coefficient $\kappa$ relates to the filling fraction $\nu$ and the Wen-Zee shift $\mathcal{S}$ via
\begin{equation}
\kappa=\nu \mathcal{S}~.
\end{equation}
Up to quadratic order and neglecting total derivative terms, the Wen-Zee term becomes
\begin{align}\label{eq:wz}
\mathcal{L}_{\text{WZ}} = \frac{\kappa}{8\pi}\left(A_0 R - \frac{1}{4\ell^2} \vareps^{ij} h_{ik} \dot h_{jk}\right)~.
\end{align}

\paragraph{Gravitational Chern-Simons term.}Another important topological contribution is the gravitational Chern-Simons (gCS) term~\cite{Witten1989,Bar-Natan1991,PhysRevLett.114.016805} appears with a quantized, dimensionless coefficient and encodes universal properties of the topological phase. Being third order in derivatives, it does not modify the dispersion of the low-energy modes. Its contribution is
\begin{align}
\begin{split}\label{eq:gcs}
\mathcal{L}_{\text{gCS}}= -\frac{\hat c}{4 \pi} \omega d \omega &= -\frac{\hat c}{4 \pi}\left( \omega_0 R -\vareps^{ij}\omega_i \dot \omega_j \right) \\
&\approx \frac{\hat c}{16\pi}\vareps^{ij}\vareps^{ab} \vareps^{cd} \d_a h_{bi} \d_c \dot h_{dj}~,    
\end{split}
\end{align}
where $\hat{c}$ is a phenomenological parameter. The coefficient $\hat c$ is fixed by anomaly inflow and equals the chiral central charge up to a conventional factor, $\hat c = c_-/12$.

Thus, the full $(2+1)$-dimensional geometric MCS Lagrangian for the chiral spin-2 graviton is
\begin{equation}\label{eq:mcs}
  \mathcal L_{\text{MCS}} = \mathcal L_{\text{Maxwell}} + \mathcal L_{\text{WZ}} + \mathcal L_{\text{gCS}}~,
\end{equation}
and, by construction, the corresponding effective action is invariant under APDs.

\subsection{Projected static structure factor}

We start from the linearized MCS Lagrangian for the chiral spin-2 mode,
\begin{equation}
\mathcal{L}[A_0,h_{ij}] = \frac{\kappa}{8\pi}\left(A_0 R - \frac{1}{4\ell^2} \vareps^{ij} h_{ik} \dot h_{jk}\right)+ c_1 \left(\d_i v_j + \d_j v_i +\dot h_{ij} \right)^2 -c_2 R^2~,   
\end{equation}
with unimodular fluctuations $h^i_i=0$ and linearized Ricci scalar $R=\partial_i\partial_j h_{ij}$. Including the linearized gravitational Chern-Simons term correction and with Weyl gauge $A_0=0$, the spin-2 Lagrangian becomes
\begin{equation}
  \mathcal L[h_{ij}]
  =-\frac{\kappa}{32\pi\ell^2}\varepsilon^{ij}h_{ik}\dot h_{jk}
   +\frac{\hat c}{16\pi}\varepsilon^{ij}\varepsilon^{ab}\varepsilon^{cd}\partial_a h_{bi}\partial_c \dot h_{dj}+c_1 \dot h_{ij}^2-c_2 R^2~.
\end{equation}

\begin{figure}[t]
  \centering
  \includegraphics[width=0.7\linewidth]{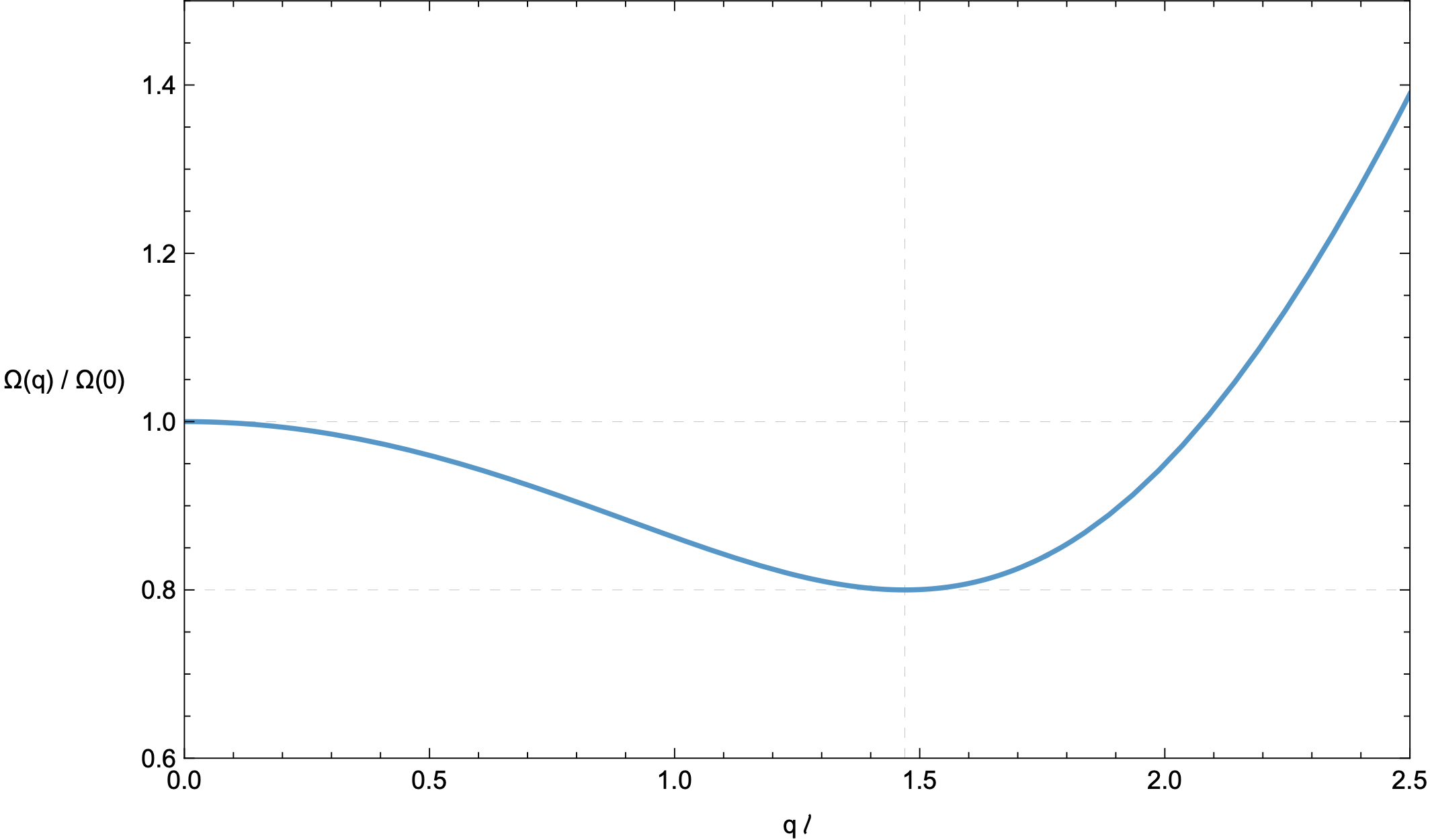}
  \caption{Dispersion relation of the collective mode at filling $\nu=1/3$ as a function of $q\ell$; only the positive physical branch is plotted. Minimum
dispersion appears at wavevector $q_{min}\ell \approx 1.47$.}
\label{fig:spin2-dispersion}
\end{figure}

To extract the dispersion, we insert a plane-wave ansatz $h_{ij}= \bar h_{ij}e^{i(\q \cdot \x -\Omega t)}$ and obtain the dispersion of the Girvin-MacDonald-Platzman (GMP) mode, see Fig.~\ref{fig:spin2-dispersion}, (a nondynamical zero mode at $\Omega=0$ is also present):
\begin{align}
    \Omega (q) = \sqrt{\frac{c_2}{2 c_1}q^4+\frac{1}{c_1^2}\left(\Omega_0+\frac{1}{2}\Omega'_0 q^2 \right)^2}~,
\end{align}
where we define
\begin{equation}
  \Omega_0\equiv-\frac{\kappa}{32\pi\ell^2}~,\qquad
  \Omega'_0\equiv\frac{\hat c}{16\pi}~,\qquad
  \tilde\Omega_0(q) \equiv \Omega_0 + \frac{1}{2} \Omega'_0 q^2~.
\end{equation}

\paragraph*{Projected static structure factor.}To probe density correlation functions, we couple an external background gauge field $\tilde{A_\mu}=(\tilde{A_0}, \tilde{A_i})$ (omitting the self and gravitational CS terms),
\begin{equation}
\begin{split}
  \mathcal L[A_0,h_{ij}, \tilde{A_0}, \tilde{A_i}]
  &=\mathcal L[A_0,h_{ij}]
   +\frac{\tilde\kappa}{8\pi}\tilde A d\omega\\
  &=\mathcal L[A_0,h_{ij}]
   +\frac{\tilde\kappa}{4\pi}\left(\varepsilon^{ij}\tilde A_0 \partial_i\omega_j
   +\varepsilon^{ij}\tilde A_i \partial_j\omega_0
   -\varepsilon^{ij}\tilde A_i\dot\omega_j\right)\\
  &=\mathcal L[A_0,h_{ij}]+\tilde A_0 \rho+\tilde A_i \, j^i_{\text{top}}~,
\end{split}
\end{equation}
which gives the density and the $U(1)$ current,
\begin{equation}
  \rho=\frac{\delta \mathcal L}{\delta \tilde A_0}
  =\frac{\tilde\kappa}{8\pi} R~,\qquad
  j^i_{\text{top}}=\frac{\delta \mathcal L}{\delta \tilde A_i}
  =\frac{\tilde\kappa}{4\pi} \varepsilon^{ij}\left(\partial_j\omega_0-\dot\omega_j\right)~.
\end{equation}

We now fix the phenomenological coefficients by matching the projected static structure factor (SSF), defined as the connected equal–time correlator of projected density operators,
\begin{equation}
  \bar s(q)=\frac{1}{\bar\rho}\int\!\frac{d\Omega}{2\pi}\,
  \big\langle \rho(\Omega, q)\,\rho(-\Omega,- q)\big\rangle_{c}~.
\end{equation}

For the SSF we also retain the probe term $(\nu/4\pi)\tilde A d\tilde A$, which renders $\bar s(q)$ unambiguous. In the LLL–projected limit the $q^2$ contribution is absent, and the small-$q$ expansion is
\begin{equation}\label{eq:s4s6}
  \bar s(q)=s_4 (q\ell)^4 + s_6 (q\ell)^6+ \mathcal O\left((q\ell)^8\right)~,
\end{equation}
with coefficients $s_4$ and $s_6$. As shown below, $s_4$ is fixed by the Wen–Zee shift~\cite{PhysRevLett.69.953}, while $s_6$ is determined by the shift together with the chiral central charge~\cite{PhysRevLett.114.016805}. The absence of $s_2$ term follows from LLL projection; without projection an $s_2$ piece generally appears.

In our theory the electron density takes the form
\begin{equation}
  \rho=\bar\rho + \frac{\tilde\kappa}{8\pi} R~,
\end{equation}
so the SSF reduces to the equal-time two-point function of the Ricci scalar:
\begin{equation}
  \bar s(q)=\left(\frac{\tilde\kappa}{8\pi}\right)^2\frac{1}{\bar\rho}
  \int\!\frac{d\Omega}{2\pi}\,
  \big\langle R(\Omega, q) R(-\Omega,-q)\big\rangle_{c}~.
\end{equation}
Evaluating the correlator with the spin-2 propagator and expressing $R$ in terms of $h_{ij}$, the density-density correlator with the gCS correction is
\begin{equation}
  \big\langle \rho(\Omega,q)\,\rho(-\Omega,-q)\big\rangle_c
  =\frac{\tilde{\kappa}^2 q^4}{128 c_1 \pi^2} \frac{i}{\Omega^2 - \frac{c_2}{2c_1}q^4- \frac{1}{c_1^2}\left(\Omega_0+\frac{1}{2}\Omega'_0 q^2 \right)^2  +i0}~.
\end{equation}

We set $\Omega_0'=0$, this gives the $q^4$ term:
\begin{equation}
  \bar s(q)=\frac{2\tilde\kappa^2}{8 \kappa \nu}(q\ell)^4~,
\end{equation}
so that $4 \kappa \nu s_4=\tilde\kappa^2$. For chiral states without Landau-level mixing the general relation is
\begin{equation}
  s_4=\frac{\mathcal S-1}{8}~.
\end{equation}
Turning on $\Omega_0'\neq 0$, we obtain the $q^6$ term with the relation $\hat{c} = 8 \nu s_6$.

\section{Stueckelberg (would-be Nambu–Goldstone) Mechanism }

The MCS sector fixes the universal small-$q$ projected structure factor and gaps the long-wavelength magnetoroton, but it does not control a spatially uniform static deformation ($\dot g_{ij}=\partial_k g_{ij}=0$). To control this homogeneous degree of freedom and align the dynamical metric $g_{ij}$ with a reference background geometry $\hat g^{\,b}_{ij}(X)$, we introduce a nonlinear, APD-invariant potential via a Stueckelberg construction. This term complements, rather than replaces, the MCS sector by providing an independent, tunable mass for the spin-2 mode. 

Within this framework, the APD symmetry-allowed mass continuously tunes the system from the isotropic phase to a nematic critical point: as the tuning parameter is varied, the gap of the chiral spin-2 excitation (the GMP precursor) softens and closes at the transition.

\subsection{Connection between two APD gauge symmetry realizations}

\paragraph*{The role of the Stueckelberg field.} The Stueckelberg trick provides a gauge-invariant mass for the spin-2 mode and may be viewed as a special case of the Higgs mechanism. We take the Stueckelberg (would-be Nambu–Goldstone) field $\phi$ to be an element of the APD Lie algebra. The associated APD group element is defined as
\begin{equation}
U(\phi)\equiv e^{\phi}~,
\end{equation}
with its action on any function $f$ is the Poisson exponential
\begin{equation}\label{eq:left-action}
  U(\phi)\,f = f + \{\phi,f\} + \frac{1}{2}\{\phi,\{\phi,f\}\} + \frac{1}{3!}\{\phi,\{\phi,\{\phi,f\}\}\}+ \cdots~.
\end{equation}

The Stueckelberg field $\phi$ transforms under two types of group actions. Under a left group action with left group parameter $\lambda$,
\begin{equation}
   U(\phi) \ \longrightarrow\ U(\phi') \quad\text{with}\quad U(\phi') = U(\lambda) \cdot U(\phi)~.
\end{equation}
Using the Baker–Campbell–Hausdorff formula,
\begin{equation}
    e^{\phi}\  \to \ e^{\phi'} = e^\lambda \cdot e^\phi = e^{\lambda +\phi +\frac{1}{2}\{\lambda, \, \phi \}+\cdots}~,
\end{equation}
which results in the nonlinear transformation of the Stueckelberg field,
\begin{equation}
    \phi \ \to \ \phi'= \lambda + \phi +\frac{1}{2} \ell^2 \varepsilon^{ij}\d_i \lambda \d_j \phi +\cdots~.
\end{equation}
To first order in $\lambda$ this leads to
\begin{equation}
 \delta^{\mathrm L}_\lambda \phi = \lambda + \frac12\{\lambda,\phi\} + \mathcal O(\lambda^2)~,   
\end{equation}
where $\delta^{\mathrm L}_\lambda$ denotes the left infinitesimal APD transformations.

\paragraph*{Maurer-Cartan form and covariant coordinate.} Introduce the right Maurer–Cartan 1–form
\begin{equation}
\mathcal{W}_\mu(\phi) \equiv (\d_\mu e^{\phi})\,e^{-\phi}~,
\end{equation}
and its expansion is given by
\begin{equation}
\mathcal{W}_\mu(\phi) = \d_\mu\phi + \frac{1}{2}\{\phi,\d_\mu\phi\}+ \frac{1}{12}\{\phi,\{\phi,\,\d_\mu\phi\}\}+ \cdots~.
\end{equation}
Given the APD gauge connection $\mathcal{A}_\beta$ (transforming as in Eq.~\eqref{apd-gauge}),
\begin{equation}
\delta_\lambda \mathcal A_\mu
  = \d_\mu\lambda + \{\lambda,\, \mathcal A_\mu\}~,
\end{equation}
define the covariant derivative
\begin{equation}
  D_\mu\phi \equiv \mathcal{W}_\mu(\phi) - \mathcal A_\mu = \d_\mu\phi+ \frac{1}{2}\{\phi,\,\d_\mu\phi\} + \frac{1}{12}\{\phi,\, \{\phi,\, \d_\mu\phi\}\}+ \cdots - \mathcal A_\mu~.
\end{equation}
Under a left APD transformation, implemented by $e^{\phi}\!\to e^{\lambda}\cdot e^{\phi}$, the Maurer–Cartan form shifts inhomogeneously,
\begin{equation}
  \delta_\lambda \big[(\d_\mu e^{\phi})\, e^{-\phi}\big]
  = \d_\mu\lambda + \{\lambda,\, (\d_\mu e^{\phi})\, e^{-\phi}\}~,
  \qquad
\end{equation}
so that $D_\mu\phi$ transforms covariantly,
\begin{equation}
\delta_\lambda D_\mu \phi = \{\lambda,\, D_\mu \phi\}~.
\end{equation}

We then define the ``covariant coordinate'',
\begin{equation}\label{eq:covX}
  X^\alpha[\phi,\mathcal A] = x^\alpha + \ell^2 \vareps^{\alpha\beta}\, D_\beta\phi~, \qquad \delta_\lambda X^\alpha = \{\lambda,\, X^\alpha\}~,
\end{equation}
which transforms like a scalar under the APD transformation.

\paragraph*{Dynamical and background geometry.} We parameterize the dynamical spatial metric in terms of a vielbein $e_i^\alpha$,
\begin{equation}
g_{ij} = \delta_{\alpha \beta} \, e_i^\alpha \, e_j^\beta~, 
\end{equation}
with spatial indices $i,j=1,2$ and internal (flat) indices $\alpha,\beta=1,2$.
The vielbein plays a crucial role in relating the internal frame and the spatial manifold and will be used to compare APD–covariant structures built from $g_{ij}$ and from the Stuckelberg sector.

Given the covariant coordinate $X^{\alpha}[\phi,\mathcal{A}]$, we define the background reference metric
\begin{equation}\label{eq:gb}
\hat g^{\,b}_{ij}(X) = \delta_{\alpha\beta}\,\partial_i X^\alpha\,\partial_j X^\beta~,
\end{equation}
the APD gauge connection $\mathcal{A}_\mu$ has a natural extension to curved spatial backgrounds; see Appendix~\ref{sec:apd-curved} for the formulation. Because $X^{\alpha}$ transforms as a scalar under APDs, the induced metric transforms tensorially,
\begin{equation}
  \delta_\lambda \hat g^{\,b}_{ij} = -\mathcal{L}_{\xi}\,\hat g^{\,b}_{ij}~, 
  \qquad \xi^i(\lambda)=\ell^2\varepsilon^{ij}\partial_j\lambda~,
\end{equation}
and similarly the dynamical metric transforms as a rank–two tensor,
\begin{equation}
  \delta_\lambda g_{ij} = -\mathcal{L}_{\xi}\,g_{ij}~.
\end{equation}

Introducing a covariant tensor
\begin{equation}
    G_{ij} =\hat g^{\,b}_{ij}(X) - g_{ij}~, \qquad G^i_j =\hat g^{\,b \, ik} G_{kj}~.
\end{equation}
By construction,
\begin{equation}
   \delta_\lambda G_{ij}=-\mathcal L_\xi G_{ij}~,
\end{equation}
thus $G_{ij}$ is APD–covariant and any scalar built locally from $G^i_j$, e.g. $\Tr\,G$, $\Tr\,G^2$, or $\det(\delta^i_j-G^i_j)$, is APD invariant.

\paragraph*{APD-invariant nonlinear mass potential.} The building block of the potential is
\begin{equation}
\mathcal{K}^i_j[g, \hat g^{\,b}(X)] = \delta^i_j - (\sqrt{\gamma})^i_j~, \qquad \gamma^i_j = \delta^i_j -G^i_j = \hat g^{\,b \, ik}(X) g_{kj}~,
\end{equation}
since $G_{ij}$ transforms covariantly under APDs, any scalar built from $\mathcal{K}$ is APD invariant. We then construct the nonlinear, APD-invariant, and ghost-free potential
\begin{equation}
    \mathcal{L}_{\text{pot}} = - \frac{m}{2} \left([\mathcal{K}^2] - [\mathcal{K}]^2 \right)~, \qquad [\mathcal{K}]=\Tr \mathcal{K}~,
\end{equation}
with phenomenological scale $m$ that sets the zero-momentum gap of the chiral spin-2 (GMP-precursor) mode.

\paragraph*{Quadratic expansion.} 
Expanding $\sqrt{\gamma}=I-\tfrac12 G+\mathcal{O}(G^2)$ gives
$\mathcal K=\tfrac12 G+\mathcal O(G^2)$ and hence
\begin{equation}
  \mathcal{L}_{\text{pot}}
  = -\frac{m}{2}\Big([\mathcal K^2]-[\mathcal K]^2\Big)
  = -\frac{m}{8}\Big(\Tr G^2 - (\Tr G)^2\Big)
  + \mathcal O(G^3)~.
\end{equation}

Imposing unimodular limit $\det g=\det \hat g^{\,b}=1$ implies $\Tr G=0$ (to linear order $h_{ii}=\hat h^{\,b}_{ii}=0$),
so the quadratic mass term reduces to
\begin{equation}\label{eq:quadratic mass}
  \mathcal{L}_{\text{pot}}
  = -\frac{m}{8}\,G^{i}_{j}G^{j}_{i}
  = -\frac{m}{8}\,G_{ij}G^{ij}~.
\end{equation}
Expanding around a common flat background,
\begin{equation}
    g_{ij}=\delta_{ij} + h_{ij} ~, \qquad  \hat g_{ij}^{\,b}=\delta_{ij} + \hat h_{ij}^{\,b}~, 
\end{equation} 
one has $G_{ij}=\hat h_{ij}^{\,b}- h_{ij}$ to linear order and
\begin{equation}
    \mathcal{L}^{\text{linear}}_{\text{pot}} = -\frac{m}{8} \left(h_{ij} - \hat h^{\,b}_{ij} \right)^2~,
\end{equation}
which preserves APD symmetry:
\begin{align}
\begin{split}
\delta h_{ij} = - \d_i \xi_j - \d_j \xi_i~, \qquad \delta \hat h_{ij}^b = -\d_i \xi_j - \d_j \xi_i~, \qquad \d_i \xi_i=0~.
\end{split}
\end{align} 

\paragraph*{Unitary gauge and the would-be Goldstone mode.} 
In unitary gauge $\phi=0$ the Maurer–Cartan form vanishes, $\mathcal W_\beta =0$, therefore
\begin{equation}
D_\beta\phi =-\mathcal A_\beta~,
\qquad
X^\alpha=x^\alpha-\ell^2\varepsilon^{\alpha\beta}\mathcal A_\beta~.    
\end{equation}

It is convenient to denote this gauge–fixed embedding by
\begin{equation}
  \mathcal X^\alpha \equiv x^\alpha-\ell^2\varepsilon^{\alpha\beta}\mathcal A_\beta~,
  \qquad
  \hat g^{\,b}_{ij}(\mathcal X)=\delta_{\alpha\beta}\,\d_i \mathcal X^\alpha\,\d_j \mathcal X^\beta~.
\end{equation}

\begin{figure}[t]
  \centering
  \includegraphics[width=1\linewidth]{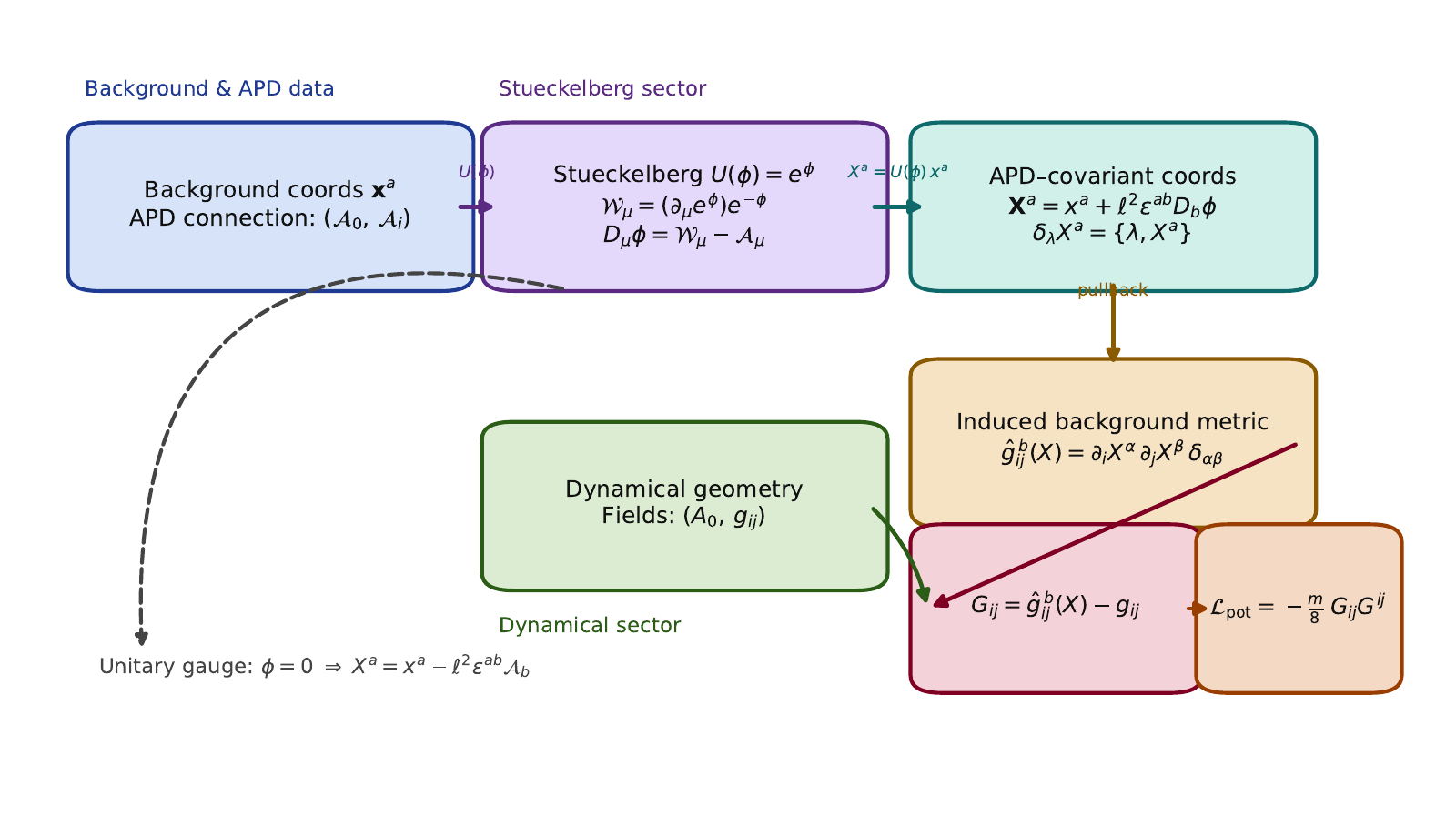}
\caption{\textbf{Role of the Stueckelberg field in the covariant–coordinate construction.}
The Stueckelberg field $\phi$ generates the APD element $U(\phi)=e^{\phi}$ and the
Maurer–Cartan form $\mathcal{W}_{\mu}=(\partial_{\mu}e^{\phi})e^{-\phi}$.
Together with the APD connection $\mathcal{A}_{\mu}$ this defines
$D_{\mu}\phi=\mathcal{W}_{\mu}-\mathcal{A}_{\mu}$ and the APD–covariant coordinates
$X^{\alpha}=x^{\alpha}+\ell^{2}\varepsilon^{\alpha\beta}D_{\beta}\phi$, which pull back the background metric to
$\hat g^{\,b}_{ij}(X)$. Comparing with the dynamical metric $g_{ij}$ yields
$G_{ij}=\hat g^{\,b}_{ij}(X)-g_{ij}$ and the APD–invariant mass potential
$\mathcal{L}_{\mathrm{pot}}=-\tfrac{m}{8}\,G_{ij}G^{ij}$.}
\label{fig:stueckelberg-covX}
\end{figure}

The Stueckelberg embedding $X^\alpha$ thus covariantizes the reference geometry, allowing a non-derivative potential without breaking APDs (Fig.~\ref{fig:stueckelberg-covX}). In unitary gauge the would-be Nambu–Goldstone mode is eaten, and the APD-invariant potential reduces (at quadratic order, in the unimodular sector) to a mass term built from
\begin{equation}
G_{ij}\equiv \hat g^{\,b}_{ij}(\mathcal X) - g_{ij}~,
\end{equation}
thereby gapping the spin-2 mode via gauge-invariant alignment of $g_{ij}$ to $\hat g^{\,b}_{ij}$. No physical global symmetry is broken: APD is a local gauge redundancy (Elitzur’s theorem), so the symmetry is Higgsed rather than explicitly broken, APD invariance remains nonlinearly realized, and no physical Goldstone survives.

\paragraph*{Nonlinear Chiral graviton theory.} Collecting terms, the effective Lagrangian 
\begin{equation}
  \mathcal L[A_0,g,g^{\,b}]
  = \frac{c_1}{4}\big(\nabla_{t} g_{ij}\big)^2 - c_2\,R^2
  - \frac{m}{2}\Big([\mathcal K^2]-[\mathcal K]^2\Big)
  + \mathcal L_{\text{top}}[A_0,g]~,
\end{equation}
where $\mathcal K$ is defined from $g_{ij}$ and $\hat g^{\,b}_{ij}$ as in the nonlinear construction, $\nabla_t$ is the covariant time derivative, $R$ is the spatial Ricci scalar of $g_{ij}$, and $\mathcal L_{\text{top}}=\mathcal L_{\text{WZ}} + \mathcal L_{\text{gCS}}$ collects the Wen–Zee~\eqref{eq:wz} and gCS~\eqref{eq:gcs} terms. At quadratic order (and in the unimodular sector) the mass potential reduces to
\begin{equation}
\mathcal L_{\text{pot}}= -\frac{m}{8}\,G_{ij}G^{ij}~,    
\end{equation}
so the parameter $m$ tunes the uniform ($q=0$) gap of the GMP spin-2 mode. The structure parallels relativistic massive gravity while remaining intrinsically nonrelativistic and not a naive nonrelativistic limit of its relativistic counterpart~\cite{PhysRevLett.106.231101}.

\subsection{Left-right (quiver) Stuckelberg construction and the mass potential}

An alternative route to the mass potential uses a left–right (quiver) Stueckelberg construction. We take two copies of the APD group, $\mathrm{APD}_L\times\mathrm{APD}_R$, acting respectively on a dynamical metric $g_{ij}(x)$ (left node) and a background metric $\hat g^{\,b}_{ij}(x)$ (right node). The link field is the APD element $U(\phi)\equiv e^{\phi}$. Under independent left/right parameters $(\lambda_L,\lambda_R)$,
\begin{equation}
  U(\phi)\ \longrightarrow\ U' \;=\; e^{\lambda_L}\,U(\phi)\,e^{-\lambda_R}~.
\end{equation}
To first order in $(\lambda_L,\lambda_R)$ (and expanding $\phi$ to linear order for clarity),
\begin{equation}
  \delta\phi = \lambda_L - \lambda_R + \frac12\{\lambda_L+\lambda_R,\phi\} + \cdots~.
\end{equation}

\begin{figure}[t]
  \centering
  \includegraphics[width=0.75\linewidth]{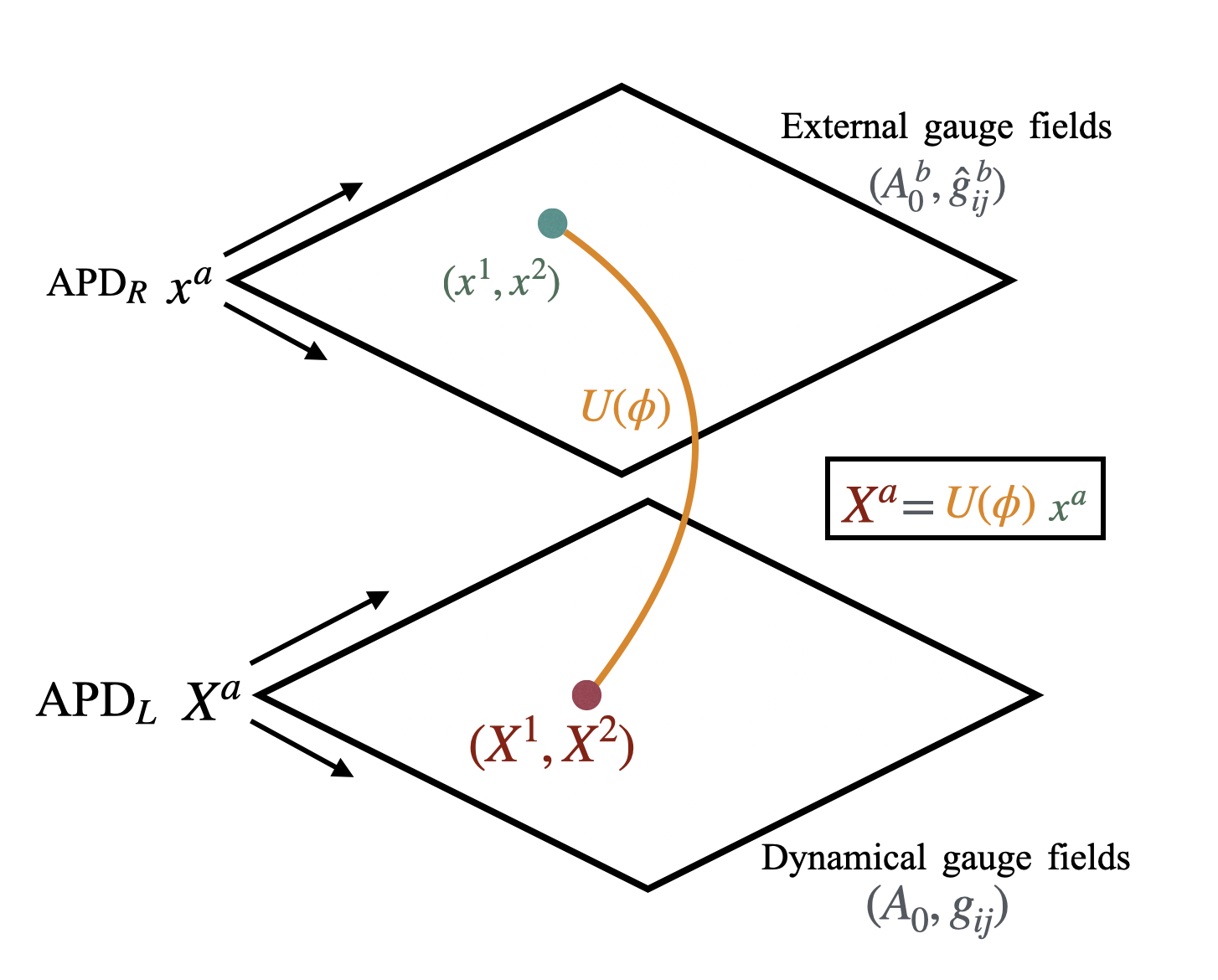}
  \caption{Stueckelberg embedding in the left–right quiver construction. The background geometry (top) carries external coordinates $x^{a}$ and gauge fields $(A^{b}_{0},\,\hat g^{\,b}_{ij})$; the dynamical geometry (bottom) carries APD–covariant coordinates $X^{a}$ and gauge fields $(A_{0},\,g_{ij})$. These two are connected by the APD group element $U(\phi)=e^{\phi}$ via $X^a=U(\phi)\!\cdot\!x^a$.}
\label{fig:goldstone-map}
\end{figure}

\paragraph*{Dressed metrics (left dynamical, right background).}
Let $\xi^i(\phi)\equiv \ell^2 \varepsilon^{ij}\partial_j\phi$ and $\mathcal L_{\xi(\phi)}$ be the Lie derivative along $\xi(\phi)$. We define a left–dressed dynamical and a right–dressed background metric by
\begin{align}
  g^{(L)}_{ij}(\phi) &\equiv e^{\mathcal L_{\xi(\phi)}}\, g_{ij}
    = g_{ij} + \mathcal L_{\xi(\phi)} g_{ij}
      + \frac12 \mathcal L_{\xi(\phi)}^2 g_{ij} + \cdots,\\
  \hat g^{\,b(R)}_{ij}(\phi) &\equiv e^{-\mathcal L_{\xi(\phi)}}\, \hat g^{\,b}_{ij}
    = \hat g^{\,b}_{ij} - \mathcal L_{\xi(\phi)} \hat g^{\,b}_{ij}
      + \frac12 \mathcal L_{\xi(\phi)}^2 \hat g^{\,b}_{ij} + \cdots~.
\end{align}
With the passive convention $\delta_\lambda f=-\mathcal L_{\xi(\lambda)}f$, these dressings transform as
\begin{equation}
\begin{split}
\delta_{\lambda_L} g_{ij}^{(L)}&=-\mathcal{L}_{\xi(\lambda_L)}\, g_{ij}^{(L)}~,\quad
\delta_{\lambda_R} g_{ij}^{(L)}=0~,\\
\delta_{\lambda_L} \hat g_{ij}^{\,b(R)}&=0~,\quad
\delta_{\lambda_R} \hat g_{ij}^{\,b(R)}=-\mathcal{L}_{\xi(\lambda_R)}\, g_{ij}^{(R)}~.
\end{split}
\end{equation}
Equivalently, introduce APD–covariant coordinates $X^a_L\!\equiv\!U(\phi)\!\cdot\!x^a$ and $X^a_R\!\equiv\!U(-\phi)\!\cdot\!x^a$ and pull back $g_{ij}$, $\hat g^{\,b}_{ij}$ with $X^a_L$, $X^a_R$ (cf. Fig.~\ref{fig:goldstone-map}). With these conventions,
\begin{equation}
  G_{ij}(\phi) \equiv \hat g^{\,b(R)}_{ij}(\phi) - g^{(L)}_{ij}(\phi)~,
\end{equation}
which transforms homogeneously under the diagonal (locked) APD, so any local scalar built from $G_{ij}$ is invariant.

\paragraph*{APD–invariant mass potentials.}A minimal unimodular quadratic potential is
\begin{equation}
\label{eq:quad}
\mathcal L_{\text{pot}}^{(2)} = -\,\frac{m}{8}\, G_{ij}(\phi)\,G^{ij}(\phi)~.
\end{equation}
Indices may be raised with $g^{(L)\,ij}$ or with a symmetric choice; around flat space these choices coincide at quadratic order. Fully nonlinear potentials can be written using $\gamma^i_j=\hat g^{b \,
ik}_{(R)}(\phi)\,g^{(L)}_{kj}(\phi)$ and $\mathcal K^i_j=\delta^i_j-(\sqrt{\gamma})^i_j$.

\paragraph*{Unitary gauge and linearization.}
In unitary gauge $\phi=0$ one has $g_{ij}^{(L)}=g_{ij}$ and $\hat g_{ij}^{\,b(R)}=\hat g_{ij}^{\,b}$, so
\begin{equation}
  G_{ij}\Big|_{\phi=0}=\hat g^{\,b}_{ij}-g_{ij}\,,
  \qquad
  \mathcal L_{\text{pot}}^{(2)}\Big|_{\phi=0}
  = -\frac{m}{8}\,(g_{ij}-\hat g^{\,b}_{ij})(g^{ij}-\hat g^{\,b\,ij})\,.
\end{equation}
Expanding about flat unimodular metrics, 
\begin{equation}
  \mathcal L_{\text{pot}}^{(2)} = -\frac{m}{8}\,\big(h_{ij}-\hat h^{\,b}_{ij}\big)^2 + \mathcal O(h^3)~,
\end{equation}
the desired APD–invariant mass term.

\paragraph*{The left and right gauge fields.}
For a purely static mass potential, no additional gauge fields are necessarily required. Alternatively, if fully local (time-dependent) APD$_L\times$APD$_R$ covariance is desired, introduce
$\mathcal A^L_\mu$ and $\mathcal A^R_\mu$ and the bi-fundamental covariant derivative
\begin{equation}
\mathcal D_\mu U \equiv \partial_\mu U - \mathcal A^L_\mu\,U + U\,\mathcal A^R_\mu~.    
\end{equation}
Setting $\mathcal A^R_\mu=0$ (and optionally $\mathcal A^L_\mu=0$) recovers the minimal one-dynamical or one-background setup.

\subsection{Road to bimetric theory: an APD–Stueckelberg perspective}

The symmetry associated with the fractional quantum Hall effect is a global symmetry that acts on the background scalar potential and the metric, both of which are traditionally treated as non-dynamical fields. One may partially gauge this symmetry by promoting the spatial metric to a dynamical field $g_{ij}$ while keeping the scalar potential $A_0$ as a background. This is the organizing idea behind bimetric descriptions of the lowest magnetoroton (spin-2) mode~\cite{PhysRevX.7.041032}: the effective action depends on both a background metric $\hat g^{\,b}_{ij}$ and an emergent dynamical metric $g_{ij}$. A potential term quadratic in the deviation between these two metrics gives the spin-2 excitation a tunable gap:
\begin{equation}\label{eq:bm}
\mathcal{L}_{\text{bm}} = -\frac{\tilde m}{2} \left(\frac{1}{2} g_{ij} \hat g^{\,b\,ij} -\gamma \right)^2~,
\end{equation}
where we assume phenomenological parameters $\ \tilde m >0$ is a phenomenological mass that sets the gap of the spin-2 mode, while $\gamma$ is a dimensionless control parameter that tunes between the isotropic and nematic phases. Because $g_{ij}$ and $\hat g^{\,b\,ij}$ are unimodular, the mixed tensor $\bar g^{i}_{j} \equiv  g^{ik}\,\hat g^{\,b}_{kj}$ has eigenvalues $(\lambda,\lambda^{-1})$, and $\frac12\,g_{ij}\,\hat g^{\,b\,ij}
=\frac12\,\mathrm{Tr}\,\bar g
=\frac12\big(\lambda+\lambda^{-1}\big)\;\ge\; 1$ with equality if and only if $g_{ij}=\hat g^{\,b}_{ij}$. It follows that for $\gamma< 1$ the potential is minimized at $g_{ij}=\hat g^{\,b}_{ij}$ (isotropic phase). In what follows we focus on
the isotropic side. For $\gamma>1$ the minimum occurs at $\frac12\,\mathrm{Tr}\,\bar g=\gamma$, which implies an anisotropy and spontaneously broken rotational symmetry (nematic phase). The phase transition is at $\gamma=1$, where the quadratic mass about the isotropic point vanishes (the spin-2 mass softens to zero). The construction of this potential term reflects the central feature of bimetric theory: a finely tuned potential that ensures the absence of ghost. Moreover, varying the effective action with respect to $A_0$ produces a charge density that satisfies the long-wavelength Girvin-MacDonald-Platzman (GMP) algebra~\cite{PhysRevB.33.2481}, a consequence of APD structure.

Comparing with the Stueckelberg construction, the quadratic APD-invariant potential~\eqref{eq:quadratic mass}:
\begin{align}\label{eq:V2}
\mathcal{L}_\text{pot} &= -\frac{m}{8} \Big(g_{ij}-\hat g^{\,b}_{ij}(X)\Big)\Big(g^{ij}-\hat g^{\,b\,ij}(X)\Big)~,
\end{align}
where $g^{\,b}_{ij}(X)$ is built from the APD scalar embedding $X^\alpha$ and hence transforms tensorially. In unitary gauge $\phi=0$, expanding about a flat, aligned background and imposing unimodularity, we write
\begin{equation}
    g_{ij}=\delta_{ij}+h_{ij}~, \qquad \hat g^{\,b}_{ij}=\delta_{ij}+\hat h^{\,b}_{ij}~, \qquad h_{ii}=\hat h^{\,b}_{ii}=0~.
\end{equation}
To quadratic order in fluctuations one finds
\begin{equation}\label{eq:mass potential}
\mathcal{L}^\text{linear}_{\text{bm}} =- \frac{\tilde m}{4}(1-\gamma)\left(h_{ij}-\hat h^{\,b}_{ij}\right)^2 + O(h^3)~, \qquad \mathcal{L}^\text{linear}_\text{pot} = - \frac{m}{8} \left(h_{ij}-\hat h^{\,b}_{ij} \right)^2 + O(h^3)~.
\end{equation}
Thus the two mass terms are equivalent at quadratic order provided\begin{equation}
  m = 2 \tilde m(1-\gamma)~.
\end{equation}

The resulting quadratic, APD-invariant bimetric Lagrangian takes the form
\begin{equation}\label{bimetric}
\mathcal{L}[h_{ij}] = -\frac{\kappa}{32 \pi \ell^2} \vareps^{ij} h_{ik} \dot h_{jk} +\frac{\hat c}{16\pi}\varepsilon^{ij}\varepsilon^{ab}\varepsilon^{cd}\partial_a h_{bi}\partial_c \dot h_{dj}+c_1 \dot h_{ij}^2 - c_2 R^2 - \frac{\tilde m}{4}(1-\gamma) \left(h_{ij} -\hat h_{ij}^{\,b} \right)^2~,
\end{equation}
where the first two parity-odd terms encode chirality (Wen–Zee and gCS), the Maxwell sector provides the parity-even kinetics, and the last term supplies the tunable mass inherited from either~\eqref{eq:bm} or~\eqref{eq:V2}.

\subsubsection{Dispersion of the neutral collective mode}

We diagonalize the quadratic theory by inserting a plane-wave ansatz $h_{ij}(t,\x)=\bar h_{ij}\,e^{i(\mathbf q\cdot \mathbf x-\Omega t)}$.
For the Lagrangian
\begin{equation}
  \mathcal L[h_{ij}]
  =-\frac{\kappa}{32\pi\ell^2}\varepsilon^{ij}h_{ik}\dot h_{jk}
   +\frac{\hat c}{16\pi}\varepsilon^{ij}\varepsilon^{ab}\varepsilon^{cd}
     \partial_a h_{b i}\partial_c \dot h_{d j}
   + c_1 \dot h_{ij}^2 - c_2 R^2 + M h_{ij}^2~,
\end{equation}
with 
\begin{equation}
M\equiv \frac{\tilde m}{4}(\gamma-1)~,\qquad  \tilde\Omega_0(q)\equiv \Omega_0+\frac12 \Omega_0' q^2~,\qquad
  \Omega_0\equiv -\frac{\kappa}{32\pi\ell^2}~,\qquad
  \Omega_0'\equiv \frac{\hat c}{16\pi}~.
\end{equation}
The dispersions of the GMP mode (positive-frequency branches) are
\begin{equation}\label{eq:disp-full}
  \Omega_\pm^2(q)=\frac{1}{4c_1^2} \left[
    2 \tilde\Omega_0(q)^2 + c_1c_2 q^4 + 4c_1 M \pm \sqrt{\big(2 \tilde\Omega_0(q)^2 + c_1c_2 q^4\big)^2 + 16 c_1 M \tilde\Omega_0(q)^2}\right]~.
\end{equation}

\begin{figure}[t]
  \centering
  \includegraphics[width=0.7\linewidth]{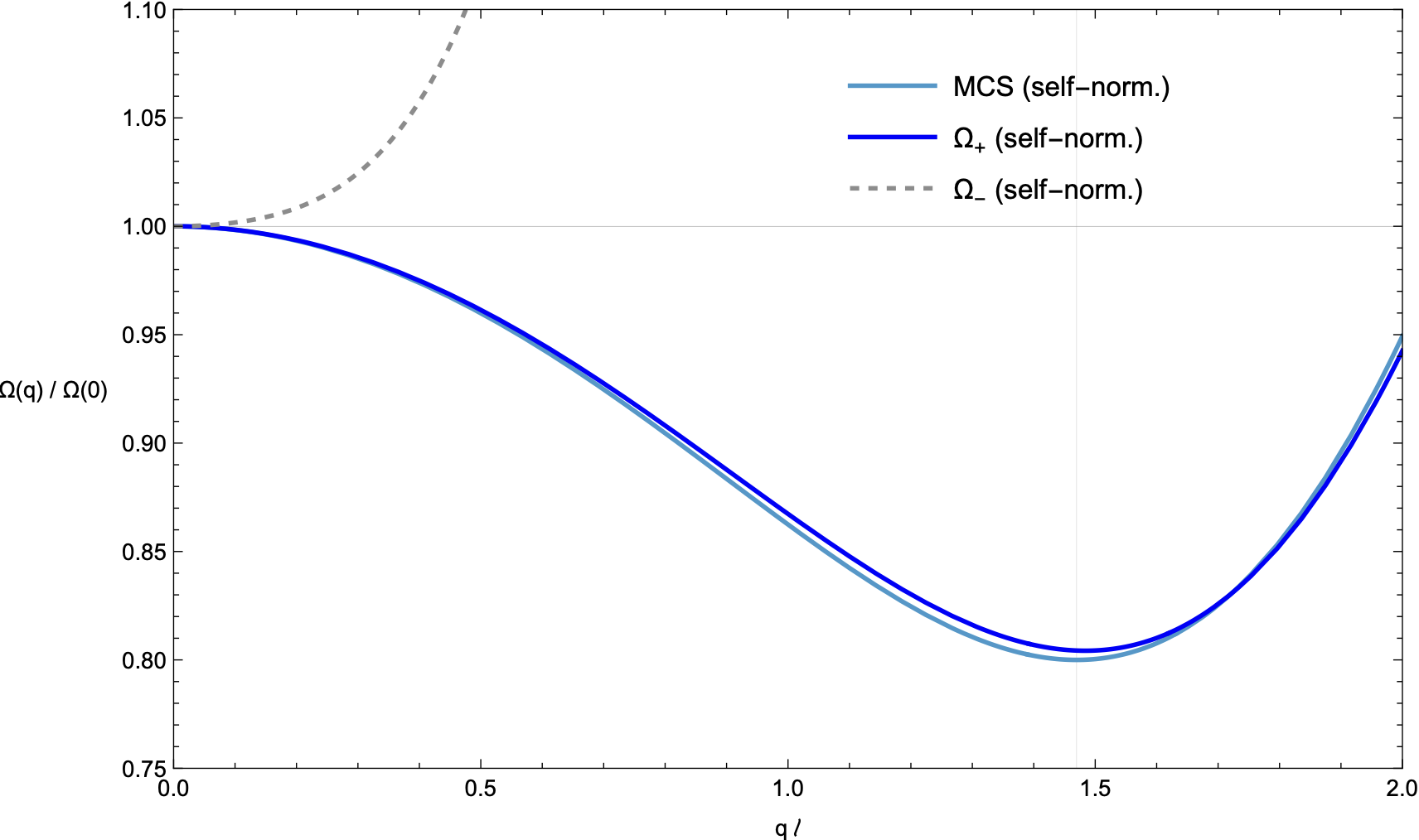}
  \caption{Collective mode dispersion relation with $\nu=1/3$ in the MCS theory augmented by an additional fine-tuning APD–Stueckelberg mass potential~\eqref{eq:disp-full}. In the limit of vanishing mass term the curve reduces continuously to the pure MCS result; increasing the mass slightly shifts the minimum. Minimum
dispersion ($\Omega_+$) appears at wavevector $q_{min}\ell \approx 1.484$.}
\label{fig:spin2-dispersion1}
\end{figure}

\begin{table}[t]
  \centering
  \renewcommand{\arraystretch}{1.2}
  \begin{tabularx}{\linewidth}{l l X}
    \toprule
    \textbf{Regime} & \textbf{Condition} & \textbf{Result (leading behavior)} \\
    \midrule
    MCS limit & $M=0$ &
    Single propagating mode:
    $\displaystyle \Omega(q)=\sqrt{\tfrac{c_2}{2c_1}q^4+\left(\tfrac{\tilde\Omega_0(q)}{c_1}\right)^2}$,
    together with a nondynamical zero mode $\Omega=0$. \\

    Small-$q$ & $q\to 0$ &
    Gaps:
    $\displaystyle \Omega_\pm^2(0)=\frac{1}{4c_1^2}\left(\sqrt{\Omega_0^2+4c_1 M} \pm |\Omega_0|\right)^2$.
    
    For $M \ll |\Omega_0|^2/c_1$:
    $\Omega_-(0)\approx M/|\Omega_0|$ and
    $\Omega_+(0)\approx |\Omega_0|/c_1 + M/|\Omega_0|$. \\

    Large $q$ & $q\ell\gg 1$ &
    $\displaystyle \Omega_+^2(q)\sim \frac{1}{2c_1^{2}}\left(\frac{\Omega_0'^2}{2}+c_1 c_2\right) q^4$,
    \quad
    $\displaystyle \Omega_-^2(q)\to \frac{M}{c_1}$.
    Thus the upper branch grows as $q^2$ (set by $c_2$ and $\Omega_0'$), while the lower branch saturates to $\sqrt{M/c_1}$. \\
    \bottomrule
  \end{tabularx}
  \caption{Asymptotic limits of the spin-2 dispersion (positive-frequency branches) from Eq.~(\ref{eq:disp-full}). Define $M\equiv \tfrac{\tilde m}{4}(\gamma-1)$.}
  \label{tab:dispersion-limits}
\end{table}
With this fine-tuning, the added Stueckelberg/bimetric mass acts as a controlled, APD-invariant perturbation that leaves universal long-wavelength data intact (Fig.~\ref{fig:spin2-dispersion1}).

\subsubsection{Advances beyond bimetric theory}
Our approach complements and extends bimetric descriptions in several ways. First, we formulate the neutral sector as a genuine gauge theory of APDs and obtain the graviton gap via a Stueckelberg (Higgs) mechanism—a local, gauge-invariant potential that aligns the dynamical unimodular metric with a reference geometry. At quadratic order this reduces to the familiar bimetric potential, but here it follows from symmetry rather than phenomenology. 

Second, we unify the connection and geometric realizations via a covariant-coordinate map, showing they are two parameterizations for the same gauge redundancy. 

Third, we separate the neutral geometric Wen–Zee and gCS terms from the compact charge sector that carries ground-state degeneracy, clarifying why APD ``Chern–Simons'' terms (built from the non-compact APD algebra) are not level-quantized and do not by themselves imply ground-state degeneracy. Because APD is non-compact, geometric CS terms here are not level-quantized. A compact, level-quantized completion may be obtained by matrix/fuzzy regularization APD $\to SU(N)$ with integer CS level; the large-$N$ limit recovers APD and connects to the $W_\infty$ higher-spin tower. Exploring this quantized completion and its higher-spin sector is left for future work.

Fourth, a parity-even Maxwell sector yields kinetics and, together with the geometric Chern-Simons terms, reproduces the universal small-$q$ constraints while providing a single tunable parameter that drives the isotropic–nematic transition by closing the $q=0$ spin-2 gap. 

Finally, by embedding band geometry into the metric, the framework applies directly to fractional Chern insulators, and it naturally generalizes to higher dimensions and to non-Abelian phases. Importantly, stability does not rely on adding Einstein–Hilbert– or ``Christoffel-connection-difference''–type kinetic terms (the former is topological in two spatial dimensions, the latter is redundant at the quadratic level in our power counting).

\subsection{The phase diagram: isotropic and nematic phases}

Expanding about a flat, aligned background in the unimodular sector, the APD-invariant, quadratic mass potential from either the bimetric or Stueckelberg construction reads in Eq.~\eqref{eq:mass potential}:
\begin{equation}
    \mathcal{L} =- \frac{\tilde m}{4}(1-\gamma)\left(h_{ij}-\hat h^{\,b}_{ij}\right)^2 ~,
\end{equation}
whose coefficient flips sign at a critical point $\gamma=1$, separating two phases.  
For $\gamma<1$ the isotropic phase is stable: the minimum occurs at $g_{ij}=\hat g^{\,b}_{ij}$ and the traceless spin-2 order parameter vanishes, $\langle Q_{ij}\rangle=0$.  
For $\gamma>1$ the isotropic minimum becomes unstable (negative mass squared). In practice higher-order APD-invariant invariants (e.g.\ $\Tr Q^4$, $[\Tr Q^2]^2$) stabilize the theory and select a nematic state with $\langle Q_{ij}\rangle\neq0$, i.e.\ spontaneous breaking of continuous rotations.

\subsubsection{Order parameter and kinematics}

We parameterize the unimodular metric by a symmetric traceless field $Q_{ij}$ via the matrix exponential,
\begin{equation}
g_{ij}=\exp{Q_{ij}}~,\qquad Q_{ij}=Q_{ji}~,\qquad Q_{ii}=0~,
\end{equation}
which makes $\det g=1$ and positivity manifest while providing a faithful nonlinear realization of the nematic order parameter. For small perturbations,
\begin{equation}
  g_{ij} = \delta_{ij} + Q_{ij} + \mathcal{O}(Q^2)~,
\end{equation}
so $Q_{ij}$ itself is the spin-2 order parameter and the chiral graviton corresponds to its dynamical fluctuation $\delta Q_{ij}$.

In our construction the APD symmetry is a local gauge redundancy and therefore cannot be spontaneously broken. The Stueckelberg potential Higgses the spin-2 sector: in unitary gauge the would-be Goldstone associated with the APD redundancy is absorbed, and the spin-2 field acquires a gap while APD invariance remains nonlinearly realized. By contrast, a genuine Goldstone mode appears only when a physical global symmetry is broken.

In the isotropic, incompressible FQH regime the unimodular metric satisfies $\langle Q_{ij}\rangle=0$, equivalently $\langle g_{ij}\rangle=\delta_{ij}$, so global spatial rotations associated with nematic order remain unbroken and no Goldstone mode appears; the spin-2 excitation is a massive chiral mode. As one tunes toward a nematic instability, this gap softens and vanishes at the nematic quantum critical point. In the nematic phase, with $\langle Q_{ij}\rangle \neq 0$, rotational symmetry is spontaneously broken and the spin-2 sector splits into a gapless orientational Goldstone mode and a gapped amplitude, or Higgs, mode. On a lattice the orientational mode is typically weakly gapped, a pseudo-Goldstone generated by the discrete rotational symmetry $C_n$. The Stueckelberg term Higgses only the APD gauge sector and is fully compatible with unbroken rotations in the isotropic phase.

\begin{figure}[t]
  \centering
  \includegraphics[width=0.9\linewidth]{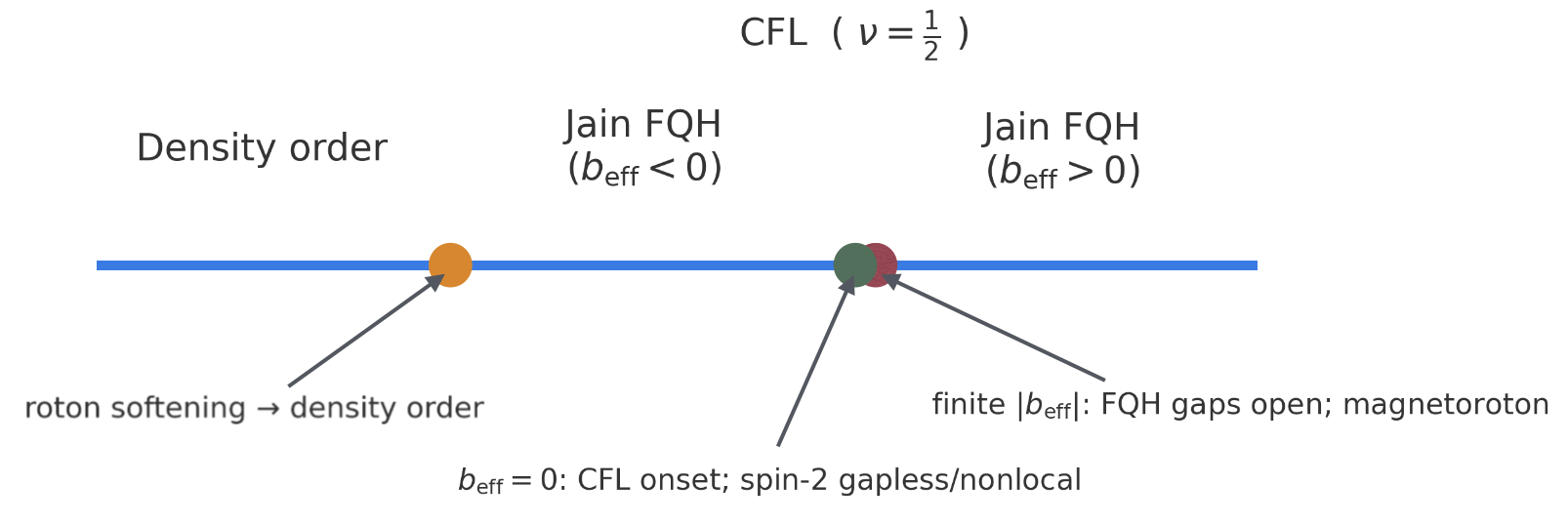}
  \caption{Schematic $T=0$ one-dimensional phase diagram along a single tuning axis:
detuning from half filling $\delta\nu$ (equivalently the composite fermion effective field $b_{\mathrm{eff}}\!\propto\! B-4\pi\rho$) at fixed microscopic couplings.
From left to right: a charge-ordered (electron-solid) phase—either a unidirectional charge-density wave (stripe) or a Wigner crystal; a gapped Jain FQH plateau in which the long-wavelength spin-2 (magnetoroton) mode is captured by a chiral graviton effective field theory with a Stueckelberg mass; the gapless composite Fermi liquid (CFL) at $b_{\mathrm{eff}}=0$, where the spin-2 sector is gapless and Landau-damped with nonlocal dynamics; and a conjugate Jain FQH plateau for the opposite sign of $b_{\mathrm{eff}}$.
Dots mark schematic transitions: (i) finite-$q$ roton softening into charge order; (ii) approach to the CFL at $b_{\mathrm{eff}}=0$ (the $q=0$ spin-2 gap closes and the mode becomes overdamped; the ``graviton'' becomes gapless without a sharp pole); (iii) FQH onset at finite $|b_{\mathrm{eff}}|$ (gaps reopen; the magnetoroton reappears).
Positions are qualitative (not to scale); the sequence and criticalities can depend on interactions, Landau-level mixing, lattice anisotropy, and disorder.}
\label{fig:phase-diagram}
\end{figure}

\subsection{Applications to fractional Chern insulators}

Fractional Chern insulators provide a natural arena for our chiral graviton framework. Upon projection to a single, isolated Chern band $\alpha$, the Bloch geometry is characterized by a momentum–space Berry curvature $\mathcal{B}(\mathbf{k})$ and the Fubini–Study (quantum) metric $g^{\alpha}_{ij}(\mathbf{k})$, which map to our dynamical metric as
\begin{equation}
\label{eq:gb-fci}
g_{ij} \equiv \frac{2}{\mathcal{B}^\alpha}\, g^{\alpha}_{ij}~,
\end{equation}
up to index conventions. Nonuniform Berry curvature and quantum metric generate symmetry allowed higher–gradient corrections and point-group anisotropies.

It is also natural to ask how explicit or spontaneous breaking of magnetic translation symmetry is encoded in this framework. A detailed quantitative implementation for the chiral graviton theory to FCIs is left for future studies.

\section{Applications of APD symmetry: Bosonization of Quantum Hall Fluids}

\subsection{Composite Fermi liquid}

The APD gauge symmetry encountered above establishes two complementary realizations.  The first, developed in Sec.~\ref{sec:gauge-symmetry}, is geometric: a chiral graviton theory formulated in terms of a dynamical unimodular spatial metric.  Here, we turn to the second, algebraic realization relevant near nematicity, in which APDs act as the kinematic algebra describing Fermi surface deformations in composite Fermi liquids. Truncating this structure to the quadrupolar sector yields a nonlinear spin-2 theory closely analogous to the unimodular chiral graviton construction; retaining the full tower leads naturally to a higher–spin framework. At long-wavelengths, quantum Hall states near half filling are well described by a Fermi liquid couple to a weak effective magnetic field. The low–energy degrees of freedom are shape deformations of the composite fermion Fermi surface, which organize systematically into an infinite tower of angular–momentum (higher–spin) harmonics~\cite{PhysRevLett.117.216403}.

Within bosonization, the spin–2 sector arises from the quadrupolar harmonics $u_{\pm2}$ of the composite fermion Fermi surface, coupled to the emergent gauge field near half filling. For Jain sequences in the isotropic, incompressible regime, these modes are gapped and generate the familiar finite–$q$ roton minimum. By contrast, in an incompressible FQH phase the long–wavelength chiral graviton is a neutral spin–2 collective mode of the guiding–center metric, captured by a chiral graviton effective theory with MCS and bimetric potential terms. In the incompressible regime the two descriptions agree at low energy: integrating out higher Fermi surface harmonics produces a gapped spin–2 sector whose coefficients (e.g.\ the projected static structure factor data $s_4,s_6$) match those of the chiral graviton theory. However, in the compressible phase at exact half filling $\nu=1/2$, the spin–2 modes hybridize with gauge fluctuations to form a gapless, Landau–damped continuum; no gapped chiral graviton mode exists, see Fig.~\ref{fig:phase-diagram}.

To construct an effective field theory for the magnetoroton within the nonlinear bosonization framework~\cite{PhysRevResearch.4.033131}, we describe multipolar deformations of the composite Fermi surface in $(2+1)$ dimensions. For fractional quantum Hall states Jain fillings at $\nu = N/(2N+1)$, composite fermions experience an effective magnetic field $b = B/(2N+1)$ and effectively form an integer quantum Hall state. In the semiclassical limit, the composite Fermi liquid corresponding to the gapless $\nu = 1/2$ state~\cite{PhysRevB.47.7312,PhysRevX.5.031027} possesses a well-defined Fermi surface with Fermi momentum related to the external magnetic field $B$ via
\begin{equation}
p_F^2 = B = \ell^{-2}~.
\end{equation}

This composite Fermi liquid state, characterized by Fermi velocity $v_F$ and Landau parameters $F_n$, occupies a region in phase space and exhibits area conservation, reflecting the property governed by the algebra of APDs. The associated kinematics is generated by fermion bilinears, the APD generators.
Following~\cite{PhysRevResearch.4.033131}, one considers the normal-ordering bilocal operator
\begin{equation}\label{bilinear}
\rho(x,y) = :\psi^\dagger(x)\psi(y):~,
\end{equation}
which closes under a magnetic deformed Poisson structure in the long-wavelength limit that captures the non-trivial phase-space geometry for composite fermions to be in a magnetic field $b$,
\begin{equation}
[\rho_\alpha,\, \rho_\beta] = \rho_{\{ \alpha, \, \beta \}_b} ~, \qquad  \{f, \, g \}_b=\d_{p^i} f \d_{x_i} g- \d_{x_i} f \d_{p^i} g- b \, \vareps_{ij} \d_{p^i} f \d_{p^j} g~,
\end{equation}
with $b=\pm B/(2N{+}1)$ for the Jain sequences. Working at frequencies and momenta of order $N^{-1}$ (relative to Fermi energy and momentum scales) gives a controlled gradient expansion in which low-energy excitations are deformations of a circular Fermi surface, encoded by a displacement field. 

\begin{figure}[t]
  \centering
  \includegraphics[width=0.4\linewidth]{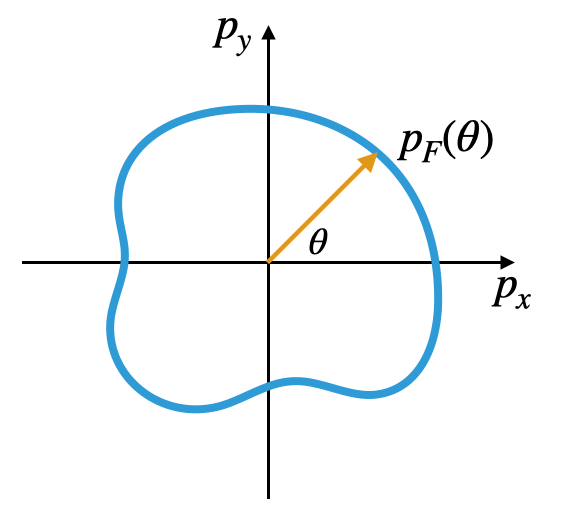}
  \caption{Deformed Fermi surface.}
\label{fig:fermi-surface}
\end{figure}

Within the bosonization framework~\cite{Haldane:1994,CastroNetoFradkin:1994,Houghton:2000bn,PhysRevB.103.165126,PhysRevResearch.4.033131}, we introduce a bosonic field $\phi(t,\x,\theta)$ expanded in angular harmonics,
\begin{equation}
\phi(t,\x,\theta)=\sum_{\ell=-\infty}^{\infty} \phi_\ell (t,\x) e^{i\ell\theta}~,
\end{equation}
and parametrize Fermi-surface dynamics by
\begin{equation}
p_F(t,\x,\theta) = p_F + \delta p_F~, \qquad  \delta p_F = u(t,\x,\theta)= \mathbf{n}_\theta \cdot \d_x \phi(t,\x,\theta)~,
\end{equation}
with the displacement field is given by
\begin{equation}
u(t,\x,\theta)\equiv \sum_{\ell=-\infty}^{\infty} u_\ell(t,\x) e^{i\ell\theta}=\sum_{\ell=-\infty}^{\infty}\left(\mathbf{n}_\theta\!\cdot\!\partial_{\mathbf{x}}\phi_\ell\right)e^{i\ell\theta}~,
\end{equation}
where $\mathbf{n}_\theta = (\cos\theta,\, \sin\theta)$ is the unit vector along the Fermi momentum direction (Fig.~\ref{fig:fermi-surface}).

\paragraph*{Nematic phase transition.}Near an isotropic–to–nematic transition, the dominant soft modes are the quadrupolar spin-2 deformations $u_{\pm 2}$, describing elliptical distortions of the Fermi surface. These modes realize the algebra of APDs, the infinite–dimensional $w_\infty$ algebra. When higher angular momentum modes are dynamically suppressed, e.g., when $\omega \gg v_F q$, the dynamics realizes a consistent spin-2 truncation, the $w_3$ subalgebra. In this limit it is natural to identify the quadrupole with the chiral graviton and to linearize the dynamical metric with unimodularity constraint and parametrize
\begin{equation}
  g_{ij}=\exp Q_{ij}~,\qquad
  Q_{ij}=\begin{pmatrix}Q_2 & Q_1\\ Q_1 & -Q_2\end{pmatrix}~,\qquad Q=Q_1+iQ_2~,
\end{equation}
where $Q$ and $\bar Q$ are complex fields. This identification underlies an effective theory of the nematic order parameter $Q_{ij}$ in quantum Hall fluids. At long-wavelengths the GMP mode is quadrupolar and serves as the precursor that softens as one approaches a continuous nematic transition.

\paragraph*{Matching effective field theory actions.}
We now match the quadratic spin–2 sector of the chiral graviton theory to Son’s Dirac composite fermion (DCF) framework~\cite{PhysRevX.5.031027}. Introduce a weakly perturbed background geometry $\hat g^{\,b}_{ij}=\delta_{ij}+\hat h^{\,b}_{ij}$ with $h^{\,b}_{ij} \ll 1$ and and $\hat h^{\,b\,i}_i=0$. 
On the chiral graviton side, the potential is minimized when the dynamical metric aligns with the background geometry 
\begin{equation}
g_{ij}=\hat g^{\,b}_{ij}~,    
\end{equation}
and we package the two independent components into $Q_1=\hat h^{\,b}_{12}$, $Q_2=\hat h^{\,b}_{11}$ and $Q\equiv Q_1+iQ_2$.

In the DCF language, the ground state is characterized by a deformed Fermi surface 
\begin{equation}
    p_F^2(\theta)= \hat g_{\,b}^{ij} p_i p_j~,
\end{equation}
with $p_i = p_F n_\theta^i$. To linear order in $\hat h^{\,b}_{ij}$ and using tracelessness $\hat h^{\,b}_{11}+ \hat h^{\,b}_{22}=0$, this gives
\begin{equation}
p_F(\theta) = p_F \left(1+\frac{\hat h^{\,b}_{11}}{2}\cos{2\theta} +\frac{\hat h^{\,b}_{12}}{2}\sin{2\theta} \right)~.
\end{equation}
Equivalently, the quadrupole harmonics of the Fermi surface deformation are
\begin{equation}
u_2 = \frac{p_F}{4}\left(\hat h^{\,b}_{11} - i \hat h^{\,b}_{12}\right)~, \qquad  u_{-2} = \frac{p_F}{4}\left(\hat h^{\,b}_{11} + i \hat h^{\,b}_{12}\right)~,
\end{equation}
with $u_{-2}=u_2^*$ at quadratic order. The linear map between the bosonized displacement modes and the chiral graviton variables is
\begin{equation}\label{linear-map}
u_2 = -\frac{i}{4} p_F Q = -\frac{i}{4 \ell} Q~, \qquad  u_{-2} = \frac{i}{4} p_F \bar Q = \frac{i}{4\ell} \bar Q~.
\end{equation}

This linear dictionary maps the quadratic, APD-invariant bimetric Lagrangian~\eqref{bimetric},
\begin{equation}
\mathcal{L}_{h_{ij}} = -\frac{\kappa}{32 \pi \ell^2} \vareps^{ij} h_{ik} \dot h_{jk} +\frac{\hat c}{16\pi}\varepsilon^{ij}\varepsilon^{ab}\varepsilon^{cd}\partial_a h_{bi}\partial_c \dot h_{dj} - \frac{\tilde m}{4}(1-\gamma) \left(h_{ij} -\hat h_{ij}^{\,b} \right)^2 +\cdots~,
\end{equation}
onto the $\ell=\pm2$ sector of the bosonized DCF theory:
\begin{equation}
\mathcal{L}_{u_{\pm 2}}= -\frac{i}{2}\frac{2N+1}{2\pi} u_2 \dot u_{-2} + \frac{i}{2}\frac{N^2 (2N+3)\ell^2}{12\pi} u_2 \Delta \dot u_{-2} - 4\ell^2 m u_2 u_{-2} + \cdots~,
\end{equation}
with the Laplace operator $\Delta=4\d \bar\d$~\footnote{We define the complex coordinates 
\begin{equation}
    z=x+iy~, \qquad \bar z=x-iy~,
\end{equation}
and derivatives: 
\begin{equation}
\partial\equiv\partial_z=\frac12(\partial_x-i\partial_y)~, \qquad \bar\partial\equiv\partial_{\bar z}=\frac12(\partial_x+i\partial_y)~. 
\end{equation}
The Laplacian is 
\begin{equation}
\Delta=\partial_i\partial_i=4 \partial\bar\partial~. 
\end{equation}
We take 
\begin{equation}
\varepsilon^{z\bar z}=-\varepsilon^{\bar z z}=2i~.
\end{equation}}.
For the Jain sequence $\nu=N/(2N+1)$ at large $N$, the universal coefficients are
$\kappa \approx (2N+1)/4$ and $\hat c \approx N^2(2N+3)/24$~\cite{PhysRevX.7.041032,Nguyen:2017qck}.
Here we are interested in the isotropic phase, tuning parameters $\gamma<1$ with $m=2\tilde m (\gamma -1)$. With these identifications, the quadratic chiral graviton theory and the bosonized DCF single-mode approximation action near the nematic phase transition agree to leading and subleading order in momentum.

\paragraph*{Further applications and outlook.}
The mapping above establishes a nontrivial correspondence between the bosonized DCF quadrupole and the quadratic chiral graviton theory, and provides a systematic procedure for coupling Fermi surface shape fluctuations to background geometry. In particular, the linearized Wen–Zee and gCS terms coincide in the two descriptions. This correspondence can be viewed as an alternative derivation of the chiral graviton effective theory. 

In realistic systems with Coulomb or pseudopotential interactions, the spin-2 mode decays into pairs of multi-roton, making it challenging to observe directly in experiments. It is an interesting open direction to explore whether the graviton can re-emerge at zero-momentum by appropriately tuning the interaction parameters to the roton-pair continuum. A systematic analysis of the nonlinear bosonization effective action, and the theory of Halperin, Lee, and Read (HLR theory)~\cite{PhysRevB.47.7312} in this language, also remains for future work.

Beyond the spin-2 truncation, retaining the full tower of Fermi surface harmonics organizes the dynamics into a nonlinear higher-spin theory governed by the quantum ($W_\infty$/GMP) deformation of APDs. This framework naturally incorporates higher-spin corrections to observables, e.g., to the projected static structure factor~\cite{Cappelli_2016}, and provides a unified view of collective excitations beyond the graviton.

\subsection{GMP/\texorpdfstring{$W_\infty$}{W∞} algebra}

As discussed in the previous sections, the nonlinear chiral graviton theory is rigidly truncated to its spin-2 sector. In FQH liquids, however, the full spectrum of shape deformations generate an infinite tower of higher-angular-momentum (higher-spin) modes. A canonical description for these deformations is provided by the algebra of APDs, i.e., the $W_\infty$ algebra. Microscopically, this algebra is realized by the commutator of the LLL projected density. For any smooth function $f(\x)$ we define the operator
\begin{equation}
    X_f = \int_\x f(\x) \rho(\x)~,
\end{equation}
where $\rho(\x)$ is the LLL projected density and serves as the generator of APDs. At long-wavelengths these operators close under the Poisson bracket,
\begin{equation}
    [X_f, \, X_g]=- i X_{\{f,\, g\}}~,\qquad   \{f, \, g\}=\ell^2 \varepsilon^{ij}\d_i f \d_j g~,
\end{equation}
which is precisely the Lie algebra of APDs, the classical $w_\infty$ algebra of guiding centers.

In momentum space, with $\rho(\mathbf{k}) \equiv X_{e^{-i\mathbf{k} \cdot \x}}$, the long-wavelength limit of the GMP algebra is
\begin{equation}
     [\rho(\mathbf{k}), \, \rho(\q)]=i \ell^2 (\mathbf{k} \times \q) \rho(\mathbf{k} + \q)~,
\end{equation}
which captures the leading algebraic structure on the LLL.

Beyond this limit, the full $W_\infty$ algebra arises as a Moyal deformation of the APD algebra. Introducing the Moyal product and bracket,
\begin{equation}
f \star g = f \exp\left( \frac i2 \ell^2 \varepsilon^{ij}\overset{\leftarrow}\partial_i    \overset{\rightarrow}\partial_j\right) g~,\qquad \{\!\{f, \, g\}\!\}= \frac{1}{i}\left[ f \star g - g \star f  \right]~,
\end{equation} 
which constitute a quantum deformation of the classical Poisson bracket. One obtains the noncommutative, associative algebra
\begin{equation}
    [X_f, \, X_g]= X_{\{\!\{f,\, g\}\!\}}~,
\end{equation}
with the operator
\begin{equation}
    X_f = \int_\x f(\x) \rho(\x)~, \qquad \rho(\mathbf{x})=:\!\psi^\dagger(\x)\psi(\x)\!:~,
\end{equation}
where $\rho(\x)$ is the normal-ordered density projected to the LLL. Equivalently, in momentum space, this structure yields the full GMP/$W_\infty$ algebra~\cite{PhysRevResearch.6.L012040},
\begin{equation}
    [\rho(\mathbf{k}), \, \rho(\q)]=2 i \exp(\ell^2 \frac{\mathbf{k} \cdot \q}{2}) \sin(\ell^2 \frac{\mathbf{k} \times \q}{2}) \rho(\mathbf{k} + \q)~.
\end{equation}

The same structure governs fractional Chern insulators (FCIs): when a Chern band has nearly uniform Berry curvature (and, more generally, up to smooth form factors), the band-projected density operators close to the GMP/$W_\infty$ algebra at long-wavelength limit~\cite{PhysRevB.85.241308,Parameswaran_2013}, providing a route to LLL physics and a natural description of the magnetoroton modes in FCIs~\cite{PhysRevB.33.2481,paul2025shininglightcollectivemodes,kousa2025theorymagnetorotonbandsmoire}. 

\subsection{Boundary theory: surface wave}

The boundary derivations in this section assume a gapped, incompressible bulk such as Laughlin, Jain, or Pfaffian states, which implies a quantized bulk Chern–Simons level and a chiral $U(1)$ Kac–Moody algebra at the edge. By contrast, the $\nu=1/2$ state on the LLL is believed to be a compressible composite Fermi liquid; the bulk is gapless and there is no protected chiral Luttinger liquid. 

At the boundaries of incompressible fractional quantum Hall fluids, low-energy excitations are described by Wen’s chiral Luttinger liquid theory~\cite{PhysRevB.41.12838,1992IJMPB...6.1711W}. The corresponding edge theory is equivalent to the Floreanini–Jackiw model~\cite{Floreanini:1987as} up to total derivatives. For multiple chiral edge modes, the general quadratic Lagrangian is
\begin{equation}\label{edge}
\mathcal{L}= \frac{K_{IJ}}{4\pi} \d_x \phi_I \d_t \phi_J - \frac{V_{IJ}}{4\pi} \d_x \phi_I \d_x \phi_J~,
\end{equation}
which yields a linear small-$q$ dispersion $\omega(q)=\frac{V}{K} q$. Here $\phi_I$ are chiral bosons; $K_{IJ}$ is the $K$-matrix characterizing the bulk topological order (e.g., for Laughlin states at filling $\nu=1/m$ one has $K=m =1/\nu$); and $V_{IJ}>0$ is a nonuniversal kernel set by the confining potential and interactions.

The emergence of these edge excitations motivates a geometric viewpoint: they are surface waves of an incompressible QH fluid, and incompressibility implies that bulk dynamics deform the shape while preserving area, i.e., they are generated by APDs, in agreement with the nonlinear bosonization framework~\cite{PhysRevResearch.4.033131}. 
The relevant symmetry is the APD group on the plane, equivalently the group of canonical transformations of the phase space generated by the Poisson bracket.

\begin{figure}[t]
  \centering
  \includegraphics[width=0.85\linewidth]{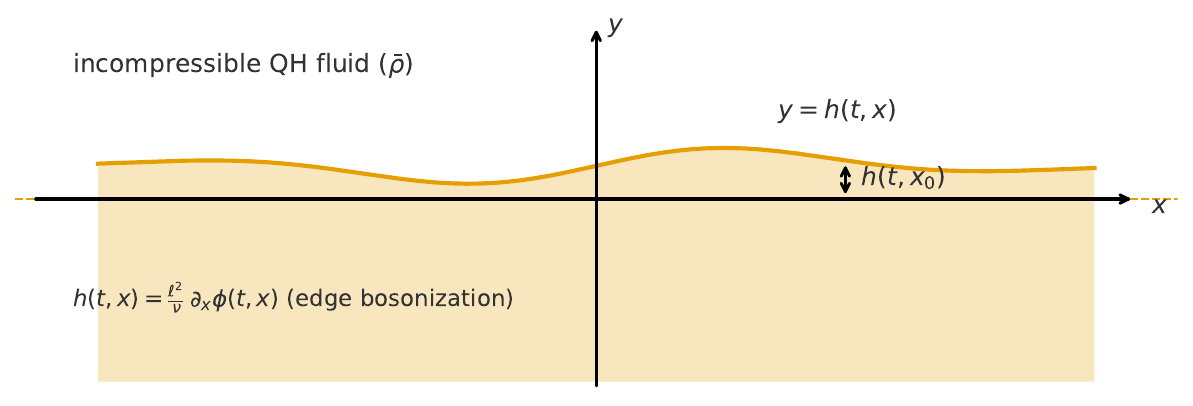}
  \caption{Incompressible quantum Hall droplet with a surface-wave boundary.
  The boundary is described by the height function $y=h(t,x)$; the droplet fills the region $y<h(t,x)$. The dashed line marks the reference boundary $y=0$, and $h(t,x_0)$ denotes a local displacement.}
  \label{fig:qh-boundary}
\end{figure}

We model a sharp, incompressible droplet of uniform bulk density $\bar\rho$ occupying the region $y<h(t,x)$, see Fig.~\ref{fig:qh-boundary}. At the equilibrium edge sits at $y=0$, therefore
\begin{equation}
\rho_0(x,y)=\bar\rho \ \Theta(-y)~.    
\end{equation}
If the edge is displaced to $y=h(t,x)$ (the local height of the boundary), the deformed profile is
\begin{equation}
    \rho(t,x,y)=\bar\rho \ \Theta\left(h(t,x)-y\right)~.
\end{equation}
Define the excess density near the edge by
\begin{equation}
  \delta\rho(t,x,y)\equiv \rho(t,x,y)-\rho_0(x,y)~.
\end{equation}
For small deformation $|h|\ll 1$, 
\begin{equation}
\Theta(h-y)=\Theta(-y)+h\,\partial_h\Theta(h-y)\big|_{h=0}+\mathcal{O}(h^2)
=\Theta(-y)+h \ \delta(-y)+\mathcal{O}(h^2)~,
\end{equation}
hence
\begin{equation}
  \delta\rho(t,x,y)=\bar\rho\Big[\Theta(h-y)-\Theta(-y)\Big] \approx \bar\rho \ h(t,x) \delta(y)~,
\end{equation}
and the one–dimensional edge density is
\begin{equation}
  \rho_{\text{edge}}(t,x)=\int dy \ \delta\rho=\bar\rho h(t,x)~.
\end{equation}
Identifying the standard edge density $\rho_{\text{edge}}=\frac{1}{2\pi}\partial_x\phi$, we obtain the map
\begin{equation}\label{eq:h_phi_map}
  h(t,x)=\frac{1}{2\pi\bar\rho} \partial_x\phi(t,x)
  =\frac{\ell^2}{\nu}\partial_x\phi(t,x)~,
\end{equation}
with $\ell$ is the magnetic length. The edge densities satisfy the $U(1)$ Kac-Moody algebra
\begin{equation}\label{eq:KM_edge}
  [\rho_{\text{edge}}(x),\,\rho_{\text{edge}}(y)]=\frac{i}{2\pi K}\partial_x\delta(x-y)~,
\end{equation}
and the unique local, metric–independent first order term whose symplectic form reproduces~\eqref{eq:KM_edge} is the ``Berry phase'' term
\begin{equation}
\mathcal L_{\text{top}} = \frac{K}{4\pi}\partial_x\phi\partial_t\phi~.
\end{equation}

\paragraph{Chiral Luttinger liquid.}We assume a locally straight edge at $y=0$ and introduce a smooth external confining potential $U(y)$. Near the edge we expand
\begin{equation}
U(y)=U(0)+U'(0) \ y+\frac12 U''(0) \ y^2+\cdots.    
\end{equation}
Working at fixed chemical potential $\mu=U(0)$, displacing the edge to $y=h(t,x)$ changes the potential energy by
\begin{equation}
\Delta E
=\bar\rho\int dx\!\int_{0}^{h(x)}\!\!dy\,\big[U(y)-\mu\big]
=\bar\rho\int dx\!\left[\frac12 U'(0)\ h^2+\frac16 U''(0)\ h^3+\cdots\right]~.    
\end{equation}
Thus, the leading local contribution to the energy is quadratic in the edge height,
\begin{equation}
\Delta E =\frac{\bar\rho \ U'(0)}{2}\int dx\,h^2 + \mathcal{O}(h^3)~.    
\end{equation}
Using~\eqref{eq:h_phi_map}, the quadratic edge Hamiltonian is
\begin{equation}
H_{\text{edge}}=\frac{V}{4\pi}\int dx\,(\partial_x\phi)^2~,\qquad
\frac{V}{4\pi}=\frac{\bar\rho U'(0)}{2}\left(\frac{\ell^2}{\nu}\right)^2 
\Longrightarrow
V=\frac{\ell^2}{\nu}U'(0)~.    
\end{equation}
The corresponding linear edge velocity is 
\begin{equation}
v=\frac{V}{K}=\ell^2 U'(0)~,    
\end{equation}
with $K=1/\nu$ and $\ell^2=1/B$, the confining potential sets the edge electric field $E_y=-U'(0)$, yielding the drift velocity $v=- E_y/B$.

Collecting terms, the low-energy edge Lagrangian to the leading order is
\begin{equation}
\mathcal{L}_{\text{edge}} = \frac{K}{4\pi} \d_x \phi \d_t \phi +\frac{V}{4\pi}(\partial_x\phi)^2+ \cdots~,
\end{equation}
i.e., a nonlinear single-mode chiral boson theory. For multiple chiral edge modes, the generalization realizes~\eqref{edge}. At long-wavelengths the edge mode dispersion is
\begin{equation}
  \omega(q)=\frac{V}{K} q~,
\end{equation}
with long–range Coulomb interactions modify $V$ momentum dependent, $V \to V(q)$. For unscreened Coulomb one finds $\omega(q)\sim q\log(1/qa)$ as $q\to 0$, where $a$ is a short-distance cutoff; typically $a\sim \ell$.

\paragraph{Quantum Hall fluid interface.}
In the absence of an external confining potential, the boundary of the quantum Hall fluid is a self-confining interface. By contrast, ordinary quantum Hall edges are pinned by a confining potential. A self-confining interface can freely propagate; it forms when the quantum Hall liquid is held together by attractive interactions between particles. Although the low-energy excitations in both cases are surface waves and preserve APD deformations of the quantum Hall fluid, we show that the bosonic interface mode exhibits a cubic dispersion in sharp contrast to the linear dispersion of edge excitations.

If the interface carries a line tension $\sigma$, its energy equals $\sigma$ times its length. Parameterizing the interface by a height field $y=h(t,x)$:
\begin{equation}
  H_\sigma=\sigma\!\int_{\partial\Omega}\! ds
           =\sigma\int dx\ \sqrt{1+(\partial_x h)^2}~.
\end{equation}
Subtracting the straight–edge baseline $H_\sigma^{(0)}=\sigma\int dx$, the excess energy is
\begin{equation}
  \Delta H_\sigma
  =\sigma\int dx\ \Big[\sqrt{1+(\partial_x h)^2}-1\Big]
  =\frac{\sigma}{2}\int dx \ (\partial_x h)^2
   -\frac{\sigma}{8}\int dx \ (\partial_x h)^4+\cdots~.
\end{equation}
Using the height–boson relation from Eq.~\eqref{eq:h_phi_map}, the quadratic piece becomes
\begin{equation}
  \Delta H_\sigma^{(2)}
  =\frac{\sigma}{2}\left(\frac{\ell^2}{\nu}\right)^2
   \int dx \ (\partial_x^2\phi)^2 =\frac{\Sigma}{4\pi}\int dx \ (\partial_x^2\phi)^2~,
\end{equation}
with the coefficient
\begin{equation}
\Sigma=2\pi \sigma \frac{\ell^4}{\nu^2}~.
\end{equation}
Accordingly, the self-confining interface Lagrangian acquires a higher-derivative correction,
\begin{equation}
  \Delta\mathcal{L}_{\text{edge}}
  =-\frac{\Sigma}{4\pi} (\partial_x^2\phi)^2~,
\end{equation}
and the interface mode exhibits a cubic dispersion
\begin{equation}
  \omega(q)=\frac{\Sigma}{K} q^3~.
\end{equation}

\section{Chiral Graviton Theory in (3+1) and generalized dimensions}

In $(d{+}1)$ dimensions with $d > 2$ spatial dimensions, the natural generalization of APDs is the group of volume–preserving diffeomorphisms (VPDs). Infinitesimally,
$x^i \to x^i + \xi^i$ with a divergence–free vector field $\xi^i$ satisfying $\d_i\xi^i=0$.

On a simply connected domain, this constraint is solved by introducing a $(d{-}2)$-form gauge parameter $\lambda_{i_1\cdots i_{d-2}}$:
\begin{equation}
  \xi^i(\lambda)
  = \frac{\ell^d}{(d-2)!}\varepsilon^{i i_1\cdots i_{d-1}}\partial_{i_1}\lambda_{i_2\cdots i_{d-1}}~,
\end{equation}
where $\ell$ is a fixed length scale (the analogue of the magnetic length), and $\varepsilon^{i_1\cdots i_d}$ is the Levi–Civita tensor. The commutator of the VPD generators $L_{\lambda}=\xi^i(\lambda)\d_i$ form a closed algebra
\begin{align}
\begin{split}
[L_\lambda,\,L_{\lambda'}]
&= L_{[\lambda,\,\lambda']}~,
\\
[\lambda,\,\lambda']_{i_1\cdots i_{d-2}}
&= (d-1)\Big(
\xi^j(\lambda)\,\partial_{[j}\lambda'_{\,i_1\cdots i_{d-2}]}
-\xi^j(\lambda')\,\partial_{[j}\lambda_{\,i_1\cdots i_{d-2}]}
\Big)
+ \partial_{[i_1}\Lambda_{\,i_2\cdots i_{d-2}]}~,
\\
\Lambda_{i_2\cdots i_{d-2}}
&\equiv \xi^{j}(\lambda')\,\lambda_{j i_2\cdots i_{d-2}}
- \xi^{j}(\lambda)\,\lambda'_{\,j i_2\cdots i_{d-2}}~.
\end{split}
\end{align}

It is convenient to encode this algebra in terms of higher (Nambu–Poisson) brackets~\cite{PhysRevD.7.2405,Takhtajan_1994}: for example, in $d=3$,
\begin{equation}
\{f,g,h\}=\ell^3 \varepsilon^{ijk}\, \d_i f \, \d_j g \, \d_k h~,
\end{equation}
and, in general dimensions,
\begin{equation}
\{f_1,f_2,\cdots, f_d\}=\ell^d \varepsilon^{i_1 i_2 \cdots i_d}(x)\d_{i_1} f_1 \d_{i_2} f_2 \cdots \d_{i_d} f_d~.    
\end{equation}
An infinitesimal VPD acts on a scalar $f(\x)$ by
\begin{equation}
    \delta_{\lambda} f(\x) = -\xi^i \d_i f= -\ell^d\varepsilon^{i_1 i_2 \cdots i_{d}}\d_{i_2} \lambda_{i_3 \cdots i_d} = -\{f_1,\cdots, f_{d-1}, f_d \}~,
\end{equation}
which can be rewritten in terms of the appropriate Nambu–Poisson bracket. Finite VPDs are obtained by exponentiating the generators,
\begin{equation}
    U(\lambda)=\exp(L_{\lambda})~,
\end{equation}
and compose according to the Baker–Campbell–Hausdorff formula.

\subsection{Gauge symmetry in (3+1) dimensions}

There are two convenient realizations of VPDs as gauge symmetries. To extend the geometric formulation to (3+1) dimensions, we introduce a Kalb-Ramond 2–form $B_{\mu\nu}=-B_{\nu\mu}$ with its intrinsic
$1$–form gauge redundancy
\begin{equation}
  \delta_\Lambda B_{\mu\nu}=\partial_\mu\Lambda_\nu-\partial_\nu\Lambda_\mu~,
\end{equation}
and gauge–invariant field strength is the 3–form
\begin{equation}
  H_{\mu\nu\rho} \equiv \partial_\mu B_{\nu\rho}
  + \partial_\nu B_{\rho\mu}
  + \partial_\rho B_{\mu\nu}~.
\end{equation}

A convenient realization of the VPD gauge symmetry introduces $(B_{0i},\, g_{ij})$ as the dynamical gauge fields, where $B_{0i}$ is the temporal component of Kalb-Ramond field and $g_{ij}$ is the spatial metric. Let $\lambda_\mu=(\lambda_0,\lambda_i)$ serve the 1-form gauge parameters and infinitesimal transformations are 
\begin{align}
\begin{split}
\delta B_{0i} &= \dot \lambda_i -\d_i \lambda_0 - \mathcal L_\xi B_{0i} = \dot \lambda_i -\d_i \lambda_0 -\xi^k\d_k B_{0i} -B_{0k} \d_i \xi^k  \\
&= \dot \lambda_i -\d_i \lambda_0  - \ell^3\vareps^{klm}
  (\d_k B_{0i} + B_{0k} \d_i) \d_l \lambda_m~, \\
 \delta g_{ij} & = -\mathcal L_\xi g_{ij} = -\xi^k\d_k g_{ij} - g_{kj}\d_i\xi^k - g_{ik}\d_j\xi^k\\ 
  & = -\ell^3 \varepsilon^{klm}\left(\d_k g_{ij}+ g_{kj}\d_i + g_{ik} \d_j \right)\d_l\lambda_m~,
\end{split} 
\end{align}
where we rewrites the Lie derivative $\mathcal L_\xi$ using $\xi^i=\ell^3\varepsilon^{ijk}\partial_j\lambda_k$ and $\ell$ is the ``magnetic length'' $\ell$ related to $H_{ijk}=\ell^{-3}\vareps_{ijk}$~\cite{PhysRevB.109.035135}.

We define the VPD ``drift velocity''
\begin{equation}
 v^i=\ell^3\vareps^{ijk}\d_j B_{0k}~, \qquad v_i\equiv g_{ij}v^j~.
\end{equation}
A covariantized time derivative of the metric is then
\begin{equation}
  \nabla_t g_{ij} \equiv \dot g_{ij} + \nabla_i v_j + \nabla_j v_i~.
\end{equation}
For linearized theory, the gauge–invariant ``electric'' field is
\begin{align}
E_{ij}=\dot h_{ij}+ \d_i v_j +\d_j v_i =\dot h_{ij}+\left(\varepsilon_{jkl}\partial_i+\varepsilon_{ikl}\partial_j \right)\partial_k B_{0l}~.
\end{align}

\paragraph*{The effective Lagrangian.}A minimal UV Lagrangian consistent with the VPD symmetry takes the form
\begin{align}
    \mathcal{L}=c_1 E_{ij}^2 -c_2 R~,
\end{align}
with $R$ is the spatial Ricci scalar. Unlike in two spatial dimensions, the Einstein–Hilbert term $\int R$ is not topological in $d=3$ and contributes to bulk dynamics. 

Expanding to quadratic order in metric (Gaussian theory) one may take
\begin{equation}
R \approx \frac14 \d_k h_{ij} \d_k h_{ij}-\frac{1}{2} \d_k h_{ij} \d_i h_{jk}~.
\end{equation}
In the Weyl gauge $B_{0i}=0$, the quadratic Lagrangian reduces to
\begin{align}
\mathcal{L}^{(2)}=c_1\dot h_{ij}^2 -c_2 \left(\frac14 \d_k h_{ij} \d_k h_{ij}-\frac{1}{2} \d_k h_{ij} \d_i h_{jk} \right) +\alpha h_{ii}~,
\end{align}
where $c_1$, $c_2$ are constants and $\alpha$ is a Lagrange multiplier enforcing the traceless condition $h_{ii}=0$. Varying $\mathcal{L}^{(2)}$ with respect to $h_{ab}$ gives the linear equations of motion
\begin{equation}
  -2c_1\ddot h_{ab}
  + \frac{c_2}{2} \partial^2 h_{ab}
  - \frac{c_2}{2}\big(\partial_a\partial_k h_{bk}+\partial_b\partial_k h_{ak}\big)
  + \frac{c_2}{3}\delta_{ab} \partial_a\partial_k h_{ak} = 0~.
\end{equation}
For a plane wave ansatz $h_{ij}(t,\mathbf x)=\bar h_{ij}\,e^{i(\mathbf q\cdot\mathbf x-\Omega t)}$ with momentum chosen along the $z$–axis,
$\q=(0,0,q)$, the residual symmetry is $SO(2)$ rotations in the
$x$–$y$ plane. It is convenient to organize $h_{ij}$ into three
$SO(2)$ sectors:

\emph{1. Transverse shear sector} 
$\,(a,b)\in\{(1,1),(2,2),(1,2)\}$:
\begin{equation}
  2c_1 \Omega^2 h_{ab}-\frac{c_2}{2} q^2 h_{ab}=0
  \quad\Rightarrow\quad
  \Omega^2=\frac{c_2}{4c_1} q^2~.
\end{equation}
These are the propagating shear modes with linear dispersion
$\Omega=v_s q$ with $v_s^2=c_2/(4c_1)$.

\emph{2. Mixed longitudinal–transverse sector}
$\,(a,b)\in\{(1,3),(2,3),(3,1),(3,2)\}$:
\begin{equation}
  2c_1 \Omega^2 h_{ab}=0
  \quad\Rightarrow\quad
  \Omega=0 \ \ (\text{or } h_{ab}=0)~.
\end{equation}
These components are nondynamical at quadratic order.

\emph{3. Pure longitudinal sector} $\,(a,b)=(3,3)$:
\begin{equation}
  2c_1 \Omega^2 h_{33}+\frac{c_2}{6} q^2 h_{33}=0~.
\end{equation}
For $c_1,c_2>0$ this equation admits no real, propagating solution
(other than $h_{33}=0$ once constraints such as tracelessness or
incompressibility are imposed). Hence there is no longitudinal
propagating mode in the present Gaussian, incompressible theory.

In (2+1) dimensions we have shown how Wen–Zee and gCS terms organize the low-energy chiral spin-2 mode. In (3+1) dimensions the situation is more subtle: the natural parity-odd densities are the Pontryagin terms, e.g., $F \wedge F$, $R \wedge R$, as well as Nieh–Yan invariants in the presence of torsion, while mixed ``Wen–Zee–like'' terms arise via higher-dimensional descent. In addition, with a Kalb–Ramond 2-form $B$ the (3+1) dimensions theory admits BF terms, e.g., $B \wedge F$, that can generate topological masses without breaking VPDs. A systematic classification of such topological terms consistent with VPD gauge symmetry and its breaking, and their impact on the spectrum, anomalies, and boundary terms, is technically involved and lies beyond the scope of this work; we leave a detailed analysis to future study.

Notations: the symbol $\wedge$ denotes the exterior (wedge) product of differential forms. For 1-forms $A=A_\mu dx^\mu$ and $B=B_\nu dx^\nu$,
$(A\wedge B)_{\mu\nu}=A_\mu B_\nu - A_\nu B_\mu$ and $A\wedge B=-B\wedge A$. Here $A$ is a (Abelian or non-Abelian) gauge connection 1-form with field strength 
$F \equiv dA$ (Abelian) or $F \equiv dA + A\wedge A$ (non-Abelian); 
$B$ is a Kalb–Ramond 2-form gauge field with field strength $H \equiv dB$ and gauge invariance $B \to B + d\Lambda$; 
$\omega$ is the spin connection 1-form and $R \equiv d\omega + \omega \wedge \omega$ is the curvature 2-form with components $R_{\mu\nu}^{ab}$. 
The gravitational Pontryagin 4-form is $\Tr(R\wedge R)$, and the Nieh–Yan 4-form is $d(e^a \wedge T_a)-e^a \wedge e^b \wedge R_{ab}$, where $e^a$ is the vielbein and $T^a \equiv de^a+\omega^a_b\wedge e^b$ is the torsion.

\section{Non-Abelian Fractional Quantum Hall States}

In this section, we extend the effective field theory developed for the magnetoroton to encompass non-Abelian fractional quantum Hall states. These states represent some of the most exotic phases of quantum matter and offer a promising platform for fault-tolerant topological quantum computation~\cite{NAYAK1996529,Nayak_2008,KITAEV20032}.

Recent experimental progress has sharpened interest in whether polarized Raman scattering can probe the nature of the $\nu = 5/2$ state~\cite{Banerjee2018,PhysRevResearch.3.023040}, potentially accessing bulk properties. Measurements of the thermal Hall conductance at the edge of the $\nu = 5/2$ state have been reported to be consistent with the PH-Pfaffian state~\cite{PhysRevX.5.031027}, diverging from the Moore-Read Pfaffian~\cite{MOORE1991362} and anti-Pfaffian~\cite{PhysRevLett.99.236806,PhysRevLett.99.236806} states. This apparent discrepancy with numerical simulations~\cite{PhysRevLett.119.026801} has prompted several proposals, including a disorder-stabilized thermal metal phase adiabatically connected to the PH-Pfaffian state~\cite{PhysRevLett.121.026801,PhysRevB.98.045112} and the incomplete edge thermalization.

Beyond the familiar spin-2 magnetoroton, the Pfaffian and anti-Pfaffian phases support a neutral fermionic excitation with spin $3/2$. This naturally raises the question of whether a unified supergravity description (local supersymmetry) can capture both modes, identifying the graviton and gravitino with the long-wavelength limits of the magnetoroton and the neutral fermion, respectively. Earlier attempts in this direction exist, but a bulk study of gapped non-Abelian states, such as the Read–Rezayi state~\cite{PhysRevB.59.8084,PhysRevB.60.8827,PhysRevB.81.045323,PhysRevB.81.155302}, remains incomplete.

Here we propose a complementary perspective: a supersymmetric generalization of area-preserving diffeomorphisms that yields a unified description of the $\nu=5/2$ state. We present two constructions of the associated supergravity theory and analyze their gauge symmetries and underlying algebraic structures.

\subsection{Moore-Read state}

While supersymmetry (SUSY) is actively sought in high-energy experiments, there is growing evidence for its emergence in strongly correlated quantum matter systems, including fractional quantum Hall states~\cite{PhysRevLett.107.036803,PhysRevLett.108.256807,PhysRevLett.125.077601}, topological superconductors~\cite{doi:10.1126/science.1248253}, ultracold atomic gases~\cite{PhysRevA.93.033642}, and lattice models~\cite{PhysRevB.76.075103,SACHDEV20102,PhysRevB.87.165145,PhysRevLett.117.166802}. In the FQH context, an $\mathcal{N}=(1,0)$ SUSY has been proposed as an effective edge description of the Moore–Read (MR) state~\cite{MOORE1991362}. A corresponding bulk theory governed by the same supersymmetry, however, remains an open problem. The MR state is a prominent candidate for realizing non-Abelian topological order with fractionally charged excitations, appearing at filling factor $\nu = 1$ for bosons and at half-integer for fermions. Moreover, it can be viewed as a p-wave paired state of composite fermions~\cite{PhysRevLett.66.3205,PhysRevB.61.10267}, i.e., topological bound states of electrons and vortices. From the closure of nonrelativistic superalgebra, the $\mathcal{N}=(1,0)$ theory lacks a spin-1 vector field, as the supersymmetry generator satisfies $QQ=0$, and does not readily produce the desired bulk Chern–Simons structure.

Motivated by experimental and theoretical indications of a spin-$3/2$ ``gravitino'' resonance~\cite{Haldane_2021,Yang_2012,PhysRevResearch.3.023040} coexisting with the chiral spin-2 graviton~\cite{Liang:2024dbb}, we consider an $\mathcal{N}=(1,1)$ supergravity framework. Within this formalism, the gravitino $\Psi_{i\alpha}$ (a Rarita-Schwinger field) serves as the gauge field for supersymmetry; upon absorbing the Goldstone fermion(s) it acquires a finite mass, realizing a supersymmetric analogue of the Higgs mechanism.

Algebraically, supergravity can be constructed as a supersymmetric generalization of the APDs by extending nonrelativistic unimodular gravity to its supersymmetric counterpart. Let $Q_\alpha$ (with the spinor index $\alpha=1,2$) denote the supersymmetry generators, forming a Majorana spinor $Q_\alpha=(Q,\bar Q)$. Their action on the graviton is schematically
\begin{equation}
    Q_\alpha g_{ij} \sim  \Psi_{i\alpha}~.
\end{equation}
The resulting multiplet is $(g_{ij},\Psi_{i\alpha})$. We introduce a spinor covariant derivative
\begin{equation}
    D_i =\d_i + \frac{1}{4} \omega_{i}^{\xi\, ab} \gamma_{ab}~,\qquad \gamma_{ab}= \frac{1}{2} [\gamma_{a},\, \gamma_{b}]~,
\end{equation}
where $\gamma^{a}$ are gamma matrices satisfying the Dirac algebra. An infinitesimal supersymmetry transformation $\delta_\epsilon = Q_\alpha \epsilon^\alpha$ with a Majorana parameter $\epsilon^\alpha$ acts as
\begin{equation}
\begin{split}
\delta_\epsilon \omega^{\xi \, ab}_i=0~,\qquad \delta_\epsilon e^a_i =i \gamma^a \Psi^{\alpha}_i \bar\epsilon^\alpha~, \qquad \delta_\epsilon \Psi_{i\alpha} = D_i\epsilon_\alpha~.
\end{split}
\end{equation}
Imposing extended metric compatibility leads to a spin connection
\begin{equation}
    \omega^{\xi \, ab}_i =\omega_i^{ab} + K_i^{ab}~,\qquad K^{ab}_i = -\bar \Psi^{[a}\gamma^{b]}\Psi_i +\cdots~.
\end{equation}

\subsection{Super-area-preserving diffeomorphisms}
\subsubsection{Gauge symmetry and Chern-Simons term}

An alternative route is to realize supersymmetry via super–area-preserving diffeomorphisms (super-APD). We formulate this theory on a supermanifold $\mathcal{M}^{(m|n)}$ with extended superspace coordinates
\begin{equation}
    z^A \equiv (x^i,\, \xi_\alpha,\, \bar\xi^{\dot \alpha})~,
\end{equation}
where $x^i$ ($i=1,\cdots, m$) are bosonic and $\xi_\alpha$ ($\alpha=1,\cdots, n$) are Grassmann-valued fermionic coordinates. We assume a closed, nondegenerate supersymplectic 2-form,
\begin{equation}
    \Omega = dz^A \wedge dz^B \Omega_{BA}~,\qquad  d\Omega=0~,
\end{equation}
on compact, orientable, simply connected spatial manifold. This symplectic form satisfies a condition
\begin{equation}
    \mathcal{L}_\Xi \, \Omega=0~,
\end{equation}
where $\mathcal{L}_\Xi$ is the Lie derivative along a vector $\Xi^A$. This constraint is solved by
\begin{equation}
\Xi^A=\Omega^{AB}\d_B \Lambda~,
\end{equation}
with $\Lambda$ an arbitrary superfunction. The corresponding super-APD generator
\begin{equation}\label{SUSY generator}
    L_\Lambda=\Omega^{AB}\d_B \Lambda \d_A = \Xi^A\d_A~,
\end{equation} 
closes under the graded commutator into the super-APD algebra
\begin{equation}\label{super-algebra}
    [L_{\Lambda},\, L_{\Lambda'}]_\xi=L_{\{\Lambda,\, \Lambda'\}_\xi}~,\qquad \{\Lambda,\, \Lambda'\}_\xi \equiv \Lambda \stackrel{\leftarrow}{\partial}_{A} \Omega^{AB} \stackrel{\rightarrow}{\partial}_{B} \Lambda'~.
\end{equation}
Here $[\cdot,\, \cdot]_\xi$ is the graded commutator~\footnote{For homogeneous superfunctions $f,g$, the graded commutator is given by
\begin{equation}\label{graded-comm}
[f,\, g ]_\xi \equiv f g - g  f (-1)^{[f][g]}~,
\end{equation}
where $[f]=0$ for bosons and $[f]=1$ for fermions, determined by the Grassmann parity of $f$. The fermionic coordinates $\xi_\alpha$ and $\bar\xi^{\dot\alpha}$ are superspace Grassmann-valued spinors that commute with the phase-space variables.

Properties of the graded commutator: we introduce the graded bracket for functions $f$ and $g$ defined as follows 
\begin{equation}
[f_+,g_+]_\xi \equiv [f_+,g_+],\ [f_+,g_-]_\xi \equiv [f_+,g_-] ,\ [f_-,g_-]_\xi \equiv \{f_- g_-\}_+~,
\end{equation}
where $f_+$ and $g_+$ are even functions, $f_-$ and $g_-$ are odd functions,  $[f, g]$ and $\{f, g\}_+$ denote the (ordinary) commutator and anticommutator, respectively.}
and $\{\cdot,\, \cdot\}_\xi$ is the classical super-Poisson bracket,
\begin{equation}\label{superbracket}
  \{f,g\}_\xi \equiv f \stackrel{\leftarrow}{\partial}_{A} \Omega^{AB} \stackrel{\rightarrow}{\partial}_{B} g~,\qquad
  [A]=\begin{cases}
  0,& z^A\ \text{bosonic}~,\\
  1,& z^A\ \text{fermionic}~.
  \end{cases}
\end{equation}
It obeys graded antisymmetry, a graded Leibniz rule, and the graded Jacobi identity (see Appendix~\ref{app:super-poisson} for a detailed discussion). In the quantum Hall LLL guiding-center limit,
\begin{equation}
  \Omega^{ij}=\ell^2\vareps^{ij}~,\qquad
  \Omega^{\xi\bar\xi}= \Omega^{\bar\xi\xi}=0~,
\end{equation}
so only the bosonic block is nonzero with magnetic length $\ell$.

Finite super-APD transformations are generated by exponentiating the infinitesimal generators,
\begin{equation}
   U(\Lambda) = \exp(L_{\Lambda})~,
\end{equation}
forming a super Lie group. Given any scalar superfunction $f(\x,\xi,\bar\xi)$, which is a superfield, transformations under the super-APD as
\begin{equation}
  \delta_\lambda f = \{\lambda,\, f\}_\xi~.
\end{equation}
A gauge superfield $A_\mu (\x,\, \xi,\, \bar\xi)$ transforms as
\begin{equation}\label{susygauge}
    \delta_\Lambda A_\mu = \d_\mu \Lambda + \{\Lambda,\, A_\mu \}_\xi~.
\end{equation}
Accordingly, a super-APD invariant Lagrangian can then be constructed, including in particular a Chern–Simons term:
\begin{equation}
    \mathcal{L} = \frac{k}{4\pi} s\Tr\left[ \vareps^{\mu\nu\rho} \left(A_\mu \d_\nu A_\rho +\frac{1}{3}\mathcal{A}_\mu \{\mathcal{A}_\nu,\, \mathcal{A}_\rho\}_\xi \right)\right]~,
\end{equation}
which is gauge-invariant under Eq.~\eqref{susygauge}. Here, $s\Tr$ denotes the supertrace with
\begin{equation}
s\Tr(fg)= (-1)^{[f][g]} s\Tr(gf)~.
\end{equation}

\subsection{Neutral collective excitations and super-GMP algebra}

It is known that Girvin, MacDonald, and Platzman (GMP) developed a foundational theory of neutral collective excitations, known as magnetorotons, in FQH states. In the long-wavelength regime, the magnetoroton is well described as a density wave propagating over a featureless ground state within the single-mode approximation (SMA); using this approach GMP obtained the full dispersion and reliable energy gap estimates. Subsequent studies confirmed that the GMP dispersion is remarkably accurate at long-wavelengths but deviates at length scales comparable to the magnetic length.

A key feature of the GMP theory is the algebra the projected density operators: after LLL projection these operators become noncommutative and realize the APD algebra (equivalently, the $W_\infty$ algebra). This nontrivial algebra underlies the structure of the magnetoroton dispersion and is central to the GMP construction. The framework extends naturally to non-Abelian FQH phases such as the MR state, where a magnetoroton is likewise expected. Moreover, Greiter, Wen, and Wilczek~\cite{PhysRevLett.66.3205} proposed that the paired nature of the MR state implies an additional neutral collective excitation, associated with the breaking of a Cooper pair. 

These considerations motivate a supersymmetric extension of the GMP algebra capturing both the bosonic (spin-2) and fermionic (spin-$3/2$) neutral sectors at $\nu=5/2$. We formulate the problem on a superplane with coordinates ($z,\bar z | \ \xi, \bar \xi$), where $z = x + iy$ is the holomorphic bosonic coordinate and $(\xi, \bar \xi)$ are Grassmann variables. For any superfunction $f(z,\xi)$, we define the supersymmetric extended operator
\begin{equation}\label{super-ops}
    X_f = \int_{z,\, \xi} f(z,\, \xi) \, \rho(z,\, \xi)~,
\end{equation}
with $\int_{z,\,\xi}=\int d^2 z \, d^2 \xi$ and $d^2 \xi=d \xi \, d\bar \xi$. Here, $\rho(z,\, \xi)$ is the projected superdensity on the LLL, which corresponds to the generator of the super-APD Lie algebra, defined as
\begin{equation}
\rho(z,\xi)=\sum_{i=1}^{N_{e}} \delta(z-z_i) \, \delta(\xi-\xi_i)~, \qquad \delta(\xi-\xi_i)=\delta(\xi-\xi_i) \, \delta(\bar\xi-\bar\xi_i)~,
\end{equation}
where the Grassmann delta function of a complex variable $\xi$ is quadratic. In the long-wavelength limit, the operators~\eqref{super-ops} close under the super-APD algebra:
\begin{equation}
    [X_f, \, X_g]=-i X_{\{f,\, g\}_\xi}~.
\end{equation}
Given an arbitrary function on superspace as the superfield, the density
operator on a superplane, can be expressed by
\begin{equation}
    \rho(z,\, \xi)=\phi(z) +\xi \, \psi^*(z)+\bar\xi \, \psi(z)+ \xi \, \bar\xi \, \rho(z)~,
\end{equation}
identifies $\psi$ and $\psi^*$ as spin-$\frac{1}{2}$ superpartners of the bosonic density $\rho$, with
\begin{align}
    \begin{split}
        \psi(z)=\sum_{i=1}^{N_{e}} \xi_i \, \delta(z-z_i) &~,\qquad \psi^*(z)=\sum_{i=1}^{N_{e}} \bar\xi_i \, \delta(z-z_i)~,\\
        \phi(z) &=\sum_{i=1}^{N_{e}} \xi_i \, \bar\xi_i \, \delta(z-z_i) ~.
    \end{split}
\end{align}
Fourier transformation yields the projected superdensity,
\begin{equation}\label{superdensity}
    \rho(\mathbf{k},\bm{\kappa}) = \rho(\mathbf{k}) -\frac{i}{2} \bm{\kappa} \, \psi^*(\mathbf{k}) -\frac{i}{2} \bar{\bm{\kappa}} \, \psi(\mathbf{k}) + \frac{1}{4} \bm{\kappa} \, \bar{\bm{\kappa}}\, \phi(\mathbf{k})~,
\end{equation}
where $\bm{\kappa}$ is the odd momentum, obtained via Fourier transformation with respect to the fermionic coordinate $\xi$. 

In the long-wavelength limit, the projected superdensity operators~\eqref{superdensity} satisfy the component form of the super-APD algebra~\eqref{super-algebra}, i.e., the super-GMP (or supersymmetric $W_\infty$) algebra:
\begin{equation}\label{S-APD algebra}
\begin{split}
[\rho(\mathbf{k}),\, \psi(\q)]&=i \ell^2 (\mathbf{k} \times \q) \psi(\mathbf{k} + \q)~,   \ [\rho(\mathbf{k}),\, \psi^*(\q)]=i \ell^2 (\mathbf{k} \times \q) \psi^*(\mathbf{k} + \q)~,  \\
\{\psi(\mathbf{k}),\, \psi^*(\q)\}_+&=i \ell^2 (\mathbf{k} \times \q) \phi(\mathbf{k} + \q)~,   \ [\phi(\mathbf{k}),\, \rho(\q)]=i \ell^2 (\mathbf{k} \times \q) \phi(\mathbf{k} + \q)~,  \\
\{\psi(\mathbf{k}),\, \psi(\q)\}_+&=\{\psi^*(\mathbf{k}),\, \psi^*(\q)\}_+=[\phi(\mathbf{k}), \, \phi(\q)]=0~,
\end{split}
\end{equation}
where $[\cdot,\, \cdot]$ and $\{\cdot,\, \cdot\}_+$ denote, respectively, the commutator and anticommutator. Grassmann variables satisfy the standard relations
\begin{equation}
  \{\xi_i,\,\xi_j\}_+ = \{\bar\xi_i,\,\bar\xi_j\}_+ = 0~.
\end{equation}

\subsubsection{Operator quantization}

Now, we employ the graded commutator~\eqref{graded-comm} to define the supersymmetric generalization of the Moyal bracket, hereafter referred to as the super-Moyal bracket~\cite{FRADKIN1991274,etde_6647936}:
\begin{equation}\label{super-Moyal}
    \{\!\{f, \, g\}\!\}_\xi \equiv \frac{1}{i}\left[ f \star_\xi g - g \star_\xi f (-1)^{[f][g]} \right]~, \qquad f \star_\xi g = f \exp{\left[\stackrel{\leftarrow}{\partial}_{A} \Omega^{AB} \stackrel{\rightarrow}{\partial}_{B}\right]}g~, 
\end{equation} 
where $[f] = 0$ for bosons and $[f] = 1$ for fermions, determined by the Grassmann parity of $f$. The symbol $\star_\xi$ denotes the super–Moyal star product with constant supersymplectic form $\Omega^{AB}$. 

The quantized operators~\eqref{super-ops} then obey the super-Moyal-APD algebra:
\begin{equation}\label{supersymmetric-moyal}
    [X_f,\, X_g]_\xi=X_{\{\!\{f, \, g\}\!\}_\xi}~.
\end{equation}
In the LLL limit, the supersymplectic form reduces to
\begin{equation}
\Omega^{ij} = \ell^2 \vareps^{ij}~,
\qquad
\Omega^{\xi \bar{\xi}} = \Omega^{\bar{\xi} \xi} = 0~.
\end{equation}
Under this condition and the graded commmutator~\eqref{graded-comm}, the super-Moyal-APD algebra leads to the full super-$W_\infty$ (super-GMP) algebra:
\begin{equation}
\begin{split}
[\rho(\mathbf{k}),\, \psi(\q)]&=2 i \exp(\ell^2 \frac{\mathbf{k} \cdot \q}{2}) \sin(\ell^2 \frac{\mathbf{k} \times \q}{2}) \psi(\mathbf{k} + \q)~,\\
[\rho(\mathbf{k}),\, \psi^*(\q)]&=2 i \exp(\ell^2 \frac{\mathbf{k} \cdot \q}{2}) \sin(\ell^2 \frac{\mathbf{k} \times \q}{2}) \psi^*(\mathbf{k} + \q)~, \\
\{\psi(\mathbf{k}),\, \psi^*(\q)\}_+&=2 i \exp(\ell^2 \frac{\mathbf{k} \cdot \q}{2}) \sin(\ell^2 \frac{\mathbf{k} \times \q}{2}) \phi(\mathbf{k} + \q)~,\\    
[\phi(\mathbf{k}),\, \rho(\q)]&=2 i \exp(\ell^2 \frac{\mathbf{k} \cdot \q}{2}) \sin(\ell^2 \frac{\mathbf{k} \times \q}{2}) \phi(\mathbf{k} + \q)~, \\
\{\psi(\mathbf{k}),\, \psi(\q)\}_+&=\{\psi^*(\mathbf{k}),\, \psi^*(\q)\}_+=[\phi(\mathbf{k}),\, \phi(\q)]=0~.
\end{split}
\end{equation}

\section{Conclusion and outlook}

In this paper we develop a nonlinear effective theory for the gapped chiral spin-2 ``graviton,'' the long-wavelength continuation of the magnetoroton, in both Abelian and non-Abelian FQH states, and outline a (3+1)-dimensional generalization. The central organizing principle is invariance under VPDs, treated as a gauge redundancy rather than a global symmetry. In (2+1) dimensions we formulate a dynamical unimodular metric theory coupled to a scalar potential, naturally yielding a MCS structure: a parity-even Maxwell term provides the kinetic energy, while parity-odd Wen–Zee and gCS terms encode chirality and fix the universal small-$q$ coefficients of the projected static structure factor.

To gap and control the long-wavelength spin-2 mode, we introduce a Stueckelberg potential built from an APD-covariant alignment of the dynamical unimodular metric $g_{ij}$ with an induced background $\hat g^{\,b}_{ij}(X)$ defined by covariant coordinates. In unitary gauge this reduces, at quadratic order, to a local bimetric mass $-\frac{m}{8}(g_{ij}-\hat g^{\,b}_{ij})^2$. In our construction the APD gauge symmetry is Higgsed and no physical Goldstone survives; by contrast, global rotational symmetry can be spontaneously broken at a nematic phase transition. A simple bimetric potential, $-\frac{\tilde m}{2}\big(\frac12 g_{ij}\hat g^{\,b \, ij}-\gamma\big)^2$, exhibits an isotropic phase for $\gamma\le 1$ and a nematic minimum for $\gamma>1$; around the aligned background its quadratic mass matches the Stueckelberg form with $m=2\tilde m(1-\gamma)$. The resulting quadratic theory yields two chiral spin-2 GMP-precursor branches; the $q=0$ gap is tunable by $m$ and softens at the nematic quantum critical point. Our framework extends naturally to fractional Chern insulators, providing a hydrodynamic bridge from ultraviolet microscopic models to infrared physics relevant to moir\'e materials. On a lattice, the continuous magnetic translation and rotation symmetries reduce to the magnetic space group: a discrete magnetic translation group together with a lattice point group $C_n$. Consequently, the small-$q$ SSF must be built from $C_n$–invariant polynomials and naturally decomposes into point-group irreducible representations rather than being a simple ``quotient'' of the continuum coefficients.

\paragraph*{Anyons, magnetorotons, and FCIs.}
In a continuum FQH liquid, a single fractionally charged quasiparticle is nondispersive: LLL projection and continuous magnetic translations make its energy position independent. In FCIs, the lattice periodicity breaks continuous magnetic translations down to a discrete magnetic space group, allowing isolated anyons to acquire a Bloch dispersion in the magnetic Brillouin zone. By contrast, the magnetoroton is a neutral collective branch: as $q \to 0$ it is the chiral spin-2, quadrupolar guiding-center metric mode, the graviton; at finite $q\!\sim\!\mathcal O(1/\ell)$ it is well described as a neutral quasiparticle–quasihole (anyon–anti–anyon) bound state. In both FQH and FCI settings it retains a well-defined dispersing branch. Our APD–Stueckelberg framework extends naturally to FCIs by encoding band geometry effects, Berry curvature and quantum metric variations, as well as lattice periodicity. A convenient microscopic route is the periodically modulated Landau-level realization, where a weak periodic potential generates a Chern band adiabatically connected to the LLL, enabling lattice-modified magnetoroton dispersion (including moir\'e-induced minibands) to be computed within a common geometric language.

\paragraph*{Experimental relevance.}
Our framework yields polarization-resolved spectroscopic signatures for Raman and THz probes and quantitative constraints from the projected SSF. In continuum Landau-level FQH, in the absence of Landau-level mixing and disorder, the small-$q$ projected SSF is universal; the graviton gap softens controllably upon tuning toward nematicity; and the $q \to 0$ chiral spin-2 line shows circular-polarization asymmetry fixed by chirality. A roton minimum is present and its position shifts with the tunable mass parameter, while the low-$q$ SSF remains universal in this limit. In FCIs the same polarization selection holds, but SSF coefficients are not generally universal because lattice periodicity and nonuniform band geometry introduce corrections. In an ideal-geometry regime (nearly uniform Berry curvature/quantum metric), the continuum relations are approximately recovered by replacing $\ell$ with an effective magnetic length $\ell_{\text{eff}}$ defined by the Brillouin-zone–averaged Berry curvature. Lattice point-group anisotropy further turns the nematic Goldstone into a small-gap pseudo-Goldstone in moir\'e FCIs.

\paragraph*{Connection to numerics.} We provide a linear dictionary mapping quadrupolar Fermi surface deformations in composite Fermi liquid bosonization near half-filling to metric fluctuations, yielding equivalent quadratic actions once universal long-wavelength data are identified. This enables quantitative fits of the phenomenological parameters to exact diagonalization (ED) or density matrix renormalization group (DMRG) spectra, as well as to the projected SSF and roton dispersion. For FCIs, nonuniform Berry curvature and quantum metric gradients enter as controlled higher-derivative corrections.

\paragraph*{(3+1) dimensions and beyond.} We sketch an extension to (3+1) dimensions by gauging VPDs and introducing a Kalb–Ramond 2-form on a unimodular spatial background, providing a natural arena for higher-dimensional geometric Chern–Simons terms and chiral tensor modes. Motivated by filling $\nu=5/2$, where a neutral spin-$3/2$ collective mode coexists with the graviton; a super-GMP algebra points toward a unified supergravity-like effective theory. The same logic carries over to FCIs: in Chern bands that realize non-Abelian topological order, for example, paired or Ising-like phases, one naturally anticipates an analogous neutral spin-$3/2$ excitation coupled to the spin-2 sector. At the boundary of incompressible quantum Hall fluids, the APD formulation reproduces the chiral Luttinger liquid and predicts a cubic dispersion for self-confining interfaces with line tension, in contrast to the linearly dispersing edge mode pinned by an external confining potential.

\paragraph*{Outlook.} Promising future directions include polarization-resolved Raman/THz tests of selection rules and the controlled softening of the $q=0$ graviton gap across tunable nematic instabilities in FQH and moir\'e FCI platforms; quantitative comparison to ED and DMRG with numerical extraction of phenomenological parameters; systematic incorporation of lattice point-group anisotropy and band geometry variations in FCIs; and extensions to non-Abelian phases such as Pfaffian, anti-Pfaffian, and Read–Rezayi, where coupling to a spin-$3/2$ mode offers falsifiable predictions for a supergravity-inspired effective field theory. For moir\'e FCIs, natural targets are spectroscopy of finite-$q$ GMP branches, candidate spin-$3/2$ gravitino modes, and higher-spin excitations, including multiple graviton branches across fillings, together with tuning experiments that drive the $q=0$ graviton gap below the continuum without a phase transition. Overall, our APD-covariant framework provides a unified platform for theoretical, numerical, and experimental studies of gapped chiral gravitons in both FQH and FCI systems.

\acknowledgments
We thank Sal Pace, Shinsei Ryu, Xiao-Gang Wen, and Yizhi You for related discussions. We also thank Dam Thanh Son for discussions and participation in the early stage of the project and Patrick A. Lee for encouragement. This work is supported, in part, by the U.S.\ DOE Grant No.\ DE-FG02-13ER41958, and by the Simons Collaboration on Ultra-Quantum Matter, which is a grant from the Simons Foundation (651440, 651446). This work was performed in part at the Aspen Center for Physics, which is supported by a grant from the Simons Foundation (1161654, Troyer). We also thank the Kavli Institute for Theoretical Physics for hospitality during Correlated Gapless Quantum Matter program, when part of this research was carried out, supported by the National Science Foundation under Grant No.\ NSF PHY-1748958 and PHY-2309135.

\appendix

\section{APD gauge connection on curved space}\label{sec:apd-curved}

In this Appendix, we extend the connection formulation of APD gauge symmetry to curved space. On a two-dimensional unimodular Riemannian manifold, choose a spatial vielbein $e^a_i$ so that $g_{ij}=\delta_{ab}e^a_i e^b_j$ and $\det g=1$. We define the Poisson bivector
\begin{equation}
  \theta^{ij}(x)\equiv \ell^2 \varepsilon^{ab} e_a^i \, e_b^j~,
  \qquad \{f,\, g\}_\theta=\theta^{ij}\, \partial_i f \, \partial_j g~,
\end{equation}
where $\{\cdot,\, \cdot\}_\theta$ denotes the associated curved-space Poisson bracket. APDs are generated by the vector field
$\xi^i(\lambda)=\theta^{ij}\partial_j\lambda$. Adopting the passive viewpoint,
any tensor $T$ transforms as $\delta_\lambda T=-\mathcal L_\xi T$. In particular, the Poisson bivector transforms as
\begin{equation}
  \delta_\lambda \, \theta^{ij}=-\mathcal L_\xi \, \theta^{ij} =-\xi^k\partial_k\theta^{ij}+\theta^{kj}\partial_k\xi^i+\theta^{ik}\partial_k\xi^j~.
\end{equation}
In unimodular coordinates we can choose $\det g=1$ so that
$\partial_i\theta^{ij}=0$ and $\partial_i\xi^i=0$.

\paragraph*{Gauge field.}Introduce the APD gauge connection $\mathcal{A}_\mu=(\mathcal{A}_0,\, \mathcal{A}_i)$ with APD gauge transformations
\begin{equation}\label{eq:apd}
  \delta_\lambda \mathcal{A}_0=\partial_t\lambda+\{\lambda,\, \mathcal{A}_0\}_\theta~,
  \qquad
  \delta_\lambda \mathcal{A}_i=\partial_i\lambda+\{\lambda,\, \mathcal{A}_i\}_\theta~.
\end{equation}
Define the field strength
\begin{equation}
  \mathcal{F}_{\mu\nu}=\partial_\mu \mathcal{A}_\nu-\partial_\nu \mathcal{A}_\mu-\{\mathcal{A}_\mu,\, \mathcal{A}_\nu\}_\theta~.
\end{equation}
Using the Jacobi identity for $\{\cdot,\cdot\}_\theta$, the algebra closes and
$\mathcal{F}_{\mu\nu}$ transforms covariantly:
\begin{equation}
  [\delta_\lambda,\, \delta_\eta]\mathcal{A}_\mu=\delta_{\{\lambda,\, \eta\}_\theta}\mathcal{A}_\mu~,
  \qquad
  \delta_\lambda \mathcal{F}_{\mu\nu}=\{\lambda,\, \mathcal{F}_{\mu\nu}\}_\theta~.
\end{equation}

A curved–space MCS Lagrangian is
\begin{align}
  \mathcal L_{\text{MCS}}
  &= \left[\frac{\varepsilon_0}{2} g^{ij}\mathcal{F}_{0i} \mathcal{F}_{0j}
    -\frac{1}{4\mu_0} g^{ik}g^{jl}\mathcal{F}_{ij} \mathcal{F}_{kl}\right]
   +\frac{k}{4\pi}\varepsilon^{\mu\nu\rho}
     \left(\mathcal{A}_\mu\partial_\nu \mathcal{A}_\rho + \frac{1}{3} \mathcal{A}_\mu\{\mathcal{A}_\nu,\, \mathcal{A}_\rho\}_\theta \right)~,
\end{align}
which is invariant under the APD gauge transformations~\eqref{eq:apd} up to a total derivative. In a local orthonormal frame one has $\theta^{ij}\to \ell^2\varepsilon^{ij}$ and the formulas reduce to the flat space expressions used in Sec.~\ref{sec:gauge-field}.

\section{Classical Super–Poisson Bracket and Super-APD Algebra}\label{app:super-poisson}

In this Appendix, we record the conventions for the classical super–Poisson bracket on the extended superspace 
$z^A=(x^i,\, \xi_\alpha,\, \bar\xi^{\dot\alpha})$ endowed with an even supersymplectic form $\Omega_{AB}$. The bracket generates super–area-preserving 
diffeomorphisms and satisfies graded antisymmetry, the graded Leibniz rule, and the graded Jacobi identity. In the purely bosonic limit it reduces to the ordinary Poisson bracket.

\emph{Definition classical super–Poisson bracket}: Let $f$ and $g$ be two $\mathbb{Z}_2$-graded functions defined on an extended superspace coordinates $z^A=(x^i,\, \xi_\alpha,\, \bar\xi^{\dot \alpha})$. The classical super–Poisson bracket is defined by
\begin{equation}
\{f, g\}_\xi \equiv f \stackrel{\leftarrow}{\partial}_{A} \Omega^{AB} \stackrel{\rightarrow}{\partial}_{B} g~,
\qquad
[A] =
\begin{cases}
0, & x^i \ \text{bosonic}~, \\
1, & (\xi_\alpha, \bar{\xi}^{\dot \alpha}) \ \text{fermionic}~,
\end{cases}
\end{equation}
where $\Omega^{AB}$ is an even supersymplectic form, antisymmetric in the bosonic-bosonic block, symmetric in the fermionic–fermionic block, and typically vanishing in the mixed blocks.

\emph{Properties}: the super–Poisson bracket satisfies:
\begin{align}
[\{f,g\}_\xi] &= [f] + [g] \quad (\mathrm{mod}\ 2)~, \tag*{(degree)} \\[2pt]
\{f,g\}_\xi &= -(-1)^{[f][g]}\,\{g,f\}_\xi 
\tag*{(graded antisymmetry)}~, \\[2pt]
\{f,gh\}_\xi &= \{f,g\}_\xi\,h 
+ (-1)^{[f][g]}\,g\,\{f,h\}_\xi
\tag*{(graded Leibniz)}~, \\[2pt]
(-1)^{[f][h]}\,\{f,\{g,h\}_\xi\}_\xi 
&+ \text{cyclic} = 0
\tag*{(graded Jacobi)}~.
\end{align}

\bibliography{chiralg}{}

\end{document}